\documentclass[fleqn,usenatbib]{mnras}
\usepackage{newtxtext,newtxmath}

\usepackage[T1]{fontenc}

\DeclareRobustCommand{\VAN}[3]{#2}
\let\VANthebibliography\thebibliography
\def\thebibliography{\DeclareRobustCommand{\VAN}[3]{##3}\VANthebibliography}

\usepackage{amsmath}	
\usepackage{lineno}
\usepackage{graphicx}
\usepackage{subcaption}
\usepackage[T1]{fontenc} 
\usepackage{multirow}
\usepackage{bm} 
\usepackage{subcaption}  
\usepackage{hyperref}
\usepackage{orcidlink}

\usepackage[dvipsnames]{xcolor}

\newcommand{\eqn}[1]{equation~(\ref{#1})}
\newcommand{\eqns}[2]{equations~(\ref{#1}) and (\ref{#2})}

\newcommand{\secn}[1]{Section~\ref{#1}}
\newcommand{\fig}[1]{Figure~\ref{#1}}
\newcommand{\figs}[2]{Figures~\ref{#1} and \ref{#2}}

\newcommand{\tab}[1]{Table~\ref{#1}}
\newcommand{\app}[1]{Appendix~\ref{#1}}
\newcommand{\angstrom}{\text{\normalfont\AA}}

\newcommand{\Msun}{\rm{M_\odot}}
\newcommand{\LUV}{L_{\rm UV}}
\newcommand{\MUV}{M_{\rm UV}}
\newcommand{\Mcrit}{M_{\rm crit}}
\newcommand{\kappaUV} {\mathcal{K}_{{\rm UV}}}
\newcommand{\kappaUVfid} {\mathcal{K}_{{\rm UV,fid}}}
\newcommand{\etaGamma} {\eta_{\gamma\ast}}
\newcommand{\etaGammafid} {\eta_{{\rm \gamma\ast,fid}}}
\newcommand{\xiIon}{\xi_{\rm ion}}
\newcommand{\xiIonfid}{\xi_{\rm ion,fid}}
\newcommand{\hcMpc}{h^{-1} \mathrm{cMpc}}

\defcitealias{Chakraborty2024}{CC24}
\defcitealias{Chakraborty2026}{CC26}

\title[Dynamical Dark Energy Models with Negative $\Lambda$]{Can an Anti-de Sitter Vacuum in the Dark Energy Sector Explain JWST High-Redshift Galaxy and Reionization Observations?}

\author[Chakraborty et al.]{
Anirban Chakraborty$^{1}$\thanks{E-mail: anirban@ncra.tifr.res.in, anirban.chakraborty096@gmail.com}\orcidlink{0009-0003-1717-8622},
Tirthankar Roy Choudhury$^{1}$\thanks{E-mail: tirth@ncra.tifr.res.in}\orcidlink{0000-0001-7462-8587},
Anjan Ananda Sen,$^{2}$\thanks{E-mail: aasen@jmi.ac.in}\orcidlink{0000-0001-9615-4909},
and Purba Mukherjee$^{2,3}$\thanks{E-mail: pdf.pmukherjee@jmi.ac.in}\orcidlink{0000-0002-2701-5654}
\\\\
$^{1}$National Centre for Radio Astrophysics, Tata Institute of Fundamental Research, Pune University Campus, Ganeshkhind, Pune 411007, India.\\
$^{2}$Centre for Theoretical Physics, Jamia Millia Islamia, New Delhi-110025, India.\\
$^{2}$Korea Astronomy and Space Science Institute,  Daejeon 34055, Republic of Korea.
}

\date{Accepted XXX. Received YYY; in original form ZZZ}

\pubyear{\the\year{}}

\begin{document}
\label{firstpage}
\pagerange{\pageref{firstpage}--\pageref{lastpage}}
\maketitle

\begin{abstract}

The unexpectedly large abundance of UV-bright galaxies at $z>10$ discovered by the James Webb Space Telescope poses a significant challenge to the standard $\Lambda$CDM cosmology. This work tests whether modifying the cosmological background, and thereby the growth of structures, can resolve this tension without invoking significant evolution in the astrophysical properties of early galaxies. We investigate an alternative framework featuring an anti-de Sitter vacuum in the dark energy sector, which naturally arises in quantum gravity models like string theory and can enhance early structure formation. Using a self-consistent semi-analytical model that couples galaxy evolution with reionization, we confront this scenario with a wide range of observations.
We show that while a model tailored to fit the high-$z$ UV luminosity functions (UVLFs) appears promising, it is in \textit{strong tension} with cosmological constraints from the CMB and other low-redshift probes. Conversely, models within this framework that satisfy these constraints provide only a modest boost to structure formation and \textit{fail to reproduce} the observed galaxy abundances at $z>10$. Although these models remain consistent with the cosmic reionization history, we find that this class of cosmological modifications is insufficient on its own to explain the galaxy excess.
Our study underscores the importance of holistic testing for any beyond-$\Lambda$CDM proposal; apparent success in one observational regime does not guarantee overall viability. By demonstrating the limitations of a purely cosmological solution, our results strengthen the case that evolving astrophysical properties are a necessary ingredient for solving the challenge of early galaxy formation.
\end{abstract}




\section{Introduction}
\label{sec:intro}

The standard Lambda Cold Dark Matter ($\Lambda$CDM) model is the cornerstone of modern cosmology, successfully explaining a vast range of phenomena from the cosmic microwave background (CMB) to the late-time acceleration of the Universe \citep{Smoot1992, WMAP2007, WMAP2013, Planck2014, Planck2016, Planck2018, Steigman2007_BBN_Review, Cyburt2016_BBN_Review, Cooke2018_DeuteriumAbundance, Cole2005_BAO, Eisenstein2005_BAO, Aubourg2015_BAO, Alam2017_BAO, Efstathiou1990_Lambda, Riess1998, Perlmutter1999, Weinberg2013, Betoule2014_JLASNe, Scolnic2018_Pantheon1, Abbott2019_DES}. However, this paradigm is facing a significant new challenge from the early Universe. Observations with the James Webb Space Telescope (JWST) have revealed a surprising abundance of luminous galaxies at redshifts $z \gtrsim 10$ \citep{Naidu2022, Castellano2022, Finkelstein2022, Labbe2023, Atek2023, Adams2023, Bradley2023, Whitler2023, Robertson2024, Castellano2023_UVLF, Finkelstein2023_UVLF, Gonzalez2023, Donnan2023, Harikane2023, Bouwens2023, McLeod2024, Adams2024, Finkelstein2024_UVLF, Whitler2025, Gonzalez2025, Weibel2025}. The number density of these objects significantly exceeds predictions from canonical $\Lambda$CDM-based models \citep{Haslbauer2022, Lovell2023, Boylan2023}. Because the abundance of early galaxies encodes crucial information about the growth of the first cosmic structures, this discrepancy offers a powerful test of the underlying cosmological framework.

Two primary pathways have emerged to explain these observations. The first proposes modifications to astrophysical processes, including: $(i)$ an enhanced efficiency or stochasticity of star formation \citep{Dekel2023, Li2024, Qin2023, Chakraborty2024, Chakraborty2026, Mirocha2023, Mason2023, Shen2023, Sun2023, Pallottini2023, Gelli2024, Kravtsov2024}; $(ii)$ a top-heavy stellar initial mass function \citep{Inayoshi2022, Trinca2023, Harikane2023, Ventura2024, Yung2024, Hutter2025, Lu2025}; $(iii)$ minimal dust attenuation at early times \citep{Ferrara2023, Ziparo2023, Ferrara2024, Ferrara2025}; or $(iv)$ a significant contribution from accreting black holes \citep{Inayoshi2022, Pacucci2022, Hegde2024, Fujimoto2024}.

Complementing these proposals, a second pathway considers whether the tension points to a flaw in the standard cosmology itself. This perspective has motivated explorations of beyond-$\Lambda$CDM physics aimed at enhancing early structure formation. Such efforts include invoking early dark energy \citep{Shen2024, Liu2024}, exotic dark matter candidates \citep{Gong2023, Bird2024, Davari2024}, primordial black holes \citep{Liu2022, Hutsi2023, Yuan2024, Matteri2025, Ellis2025}, or alterations to the primordial power spectrum and matter transfer function \citep{Biagetti2023, Parashari2023, Hirano2024, Padmanabhan2023}. While appealing, the viability of some of these cosmological solutions remains highly debated \citep{Sabti2024, Gouttenoire2024}.

In this work, we focus on modifications to the dark energy sector that, in turn, also affect the cosmic expansion history $H(z)$. The rate of expansion sets the Hubble drag, which resists the gravitational collapse of density perturbations and therefore regulates the formation of dark matter haloes -- the sites where galaxies eventually form \citep{Peebles1980_Book, Peacock1999_Textbook, Dodelson2020_Textbook, Padmanabhan2002_TextbookVol3}. While the positive cosmological constant in $\Lambda$CDM successfully explains the observed late-time acceleration of the Universe, it is plagued by theoretical issues like the fine-tuning and coincidence problems \citep{Weinberg1989, Padmanabhan2003_Review}. Furthermore, constructing a stable, positive vacuum energy (a de Sitter vacuum) is notoriously difficult within string theory \citep{Danielsson2018, Vafa2005, Palti2019, Agrawal2018, Obied2018, Cicoli2019, Cicoli2022}. In contrast, dark energy field potentials featuring a negative minimum -- corresponding to an Anti–de Sitter (AdS) vacuum or equivalently, a negative cosmological constant -- arise more naturally in scenarios motivated by string theory and the AdS/CFT correspondence \citep{Maldacena1999}, and therefore constitute an interesting theoretically well-motivated possibility. In fact, recently   \citet{Demirtas} have shown that a small but negative cosmological constant, with an appropriate order of magnitude so as to be cosmologically relevant at the present epoch, can be obtained within certain versions of supersymmetric string theory.  Moreover, scalar field dark energy models with a negative (AdS) minima have also been studied in the context of gravitational effective field theory with holographic duals \citep{Antonini:2022fna}, where they have been shown to provide an excellent fit to cosmological observations \citep{VanRaamsdonk:2023ion}, as well as in the context of the minisuperspace model, where our Universe is merging with baby universes \citep{Muralidharan:2024hsc}. Beyond their theoretical appeal, such models also have profound implications for the finite lifetime of the Universe, as investigated recently in \citep{Luu:2025dax}. Consequently,  cosmological models featuring a composite dark energy sector with a negative cosmological constant ($\Lambda < 0$), supplemented by a dynamical scalar field ($\phi$) with positive energy density, have been increasingly advocated in the literature as attractive alternatives that can satisfy the late-time acceleration constraints as well as remain consistent with other low redshift cosmological observations \citep{Cardenas2003, Dutta2018, Visinelli2019, DiValentino2021Entrp, Akarsu2020, Calderon2021, AnjanSen2023, Adil2023, Adil2024, Wang2025PhRvD, Mukherjee2025}.

Recent studies have further shown that these models can enhance structure formation at high redshifts, making them promising candidates for explaining the JWST galaxy excess \citep{Adil2023, Menci2024, Menci2024a}. As cosmic reionization is driven by these same galaxy populations, any modification to their abundance will necessarily impact the ionization state of the intergalactic medium \citep{Wise2014, Robertson2015, Bouwens2015_sources, Dayal2020, Atek2024_Spectroscopy}.

Therefore, our primary objective is to perform a rigorous and \textit{simultaneous} test of this theoretically motivated cosmological scenario. We investigate whether a model with a negative cosmological constant can \textit{alone} reproduce the observed abundance of UV-luminous galaxies across the redshift range $5 \leq z < 14$ \textit{and} satisfy the constraints on cosmic reionization. By demanding that the solution works without any evolution in the astrophysical properties of galaxies, we subject this cosmological framework to its most stringent possible test.

This paper is organized as follows. In \secn{sec:theory_model}, we detail the theoretical framework used in this work. \secn{sec:data_and_methods} describes the observational datasets and our analysis methodology. We present and discuss our results in \secn{sec:results} and summarize our main conclusions in \secn{sec:conclusion}. Unless stated otherwise, we adopt the best-fitting cosmological parameters from the Planck 2018 analysis for the baseline $\Lambda$CDM model: $\Omega_m$ = 0.3158, $\Omega_{\mathrm{DE}}$ = 0.6841077, $\Omega_r$ = $9.23\times 10^{-5}$,  $\Omega_b$ = 0.0493, $h$ = 0.6732, $\sigma_8$ = 0.8120 and $n_s$ = 0.96605 \citep{Planck2018}.

\section{Theoretical Formalism}
\label{sec:theory_model}

\subsection{Cosmological Model: Structure Formation and Halo Statistics}
\label{subsec:cosmology_framework}

In this subsection, we describe the background cosmological model and the framework for computing the growth of matter perturbations and other related cosmological quantities, such as the halo mass function, which serves as input to the galaxy formation and evolution model discussed in the next subsection.

\subsubsection{Background Cosmological Evolution}

We assume a spatially flat, homogeneous, and isotropic Universe comprising radiation, non-relativistic matter, and dark energy. For the dark energy sector, we consider an evolving dark energy component, modelled as a rolling scalar field ($\phi$), in the presence of a cosmological constant $\Lambda$, which is negative \citep{AnjanSen2023,Adil2023, Menci2024}. We adopt one of the most widely used parameterizations, known as the Chevallier-Polarski-Linder (CPL) parameterization \citep{Chevallier2001, Linder2003}, to describe the equation of state, $w_{\phi} = P_{\phi}/\rho_{\phi}$, for the field $\phi$, which varies with redshift $z$ as  - 
\begin{eqnarray}
\label{eq:cpl_eos}
    w_{\phi}(z) = w_{0} + w_{a} \left(\frac{z}{1+z}\right) 
\end{eqnarray}
where $w_{0}$ and $w_{a}$ are constants determining the value of the equation of state and its rate of change (with respect to the scale factor) at the present epoch, respectively. As per the above parameterization, the equation of state smoothly evolves from a value of $w_{\phi}  = w_0$ at $z = 0$ to $w_{\phi} \rightarrow w_0 + w_a$ as $z \rightarrow \infty$. 

Assuming a spatially flat Universe, the first Friedmann equation, which describes the evolution of the Hubble expansion rate, can be written as - 
\begin{multline}
H^2(z) = H_0^2~\Bigg[ \Omega_r (1+z)^4 + \Omega_m(1+z)^3 + \Omega_{\Lambda} \\
+ \Omega_\phi (1+z)^{3(1+w_0+w_a)} \exp \Bigg( -3 w_a \frac{z}{1+z} \Bigg) \Bigg]
\label{eq:hubble_expansion_rate}
\end{multline}

where $H_0$ is the Hubble constant and $\Omega_i$ is the present-day density parameter of the $i$-th component, such as radiation, matter, a cosmological constant, and the scalar field $\phi$.

From \eqn{eq:hubble_expansion_rate}, it is clear that larger values of $w_0$ and $w_a$ generally lead to greater dark energy density at higher redshifts, and result in a higher rate of cosmic expansion $H(z)$. 

We identify the total dark energy (DE) sector as the combination of two components: the CPL scalar field, characterised by the density parameter  $\Omega_\phi$, and the cosmological constant, characterised by the density parameter $\Omega_\Lambda$. Consequently, it immediately follows that 
\begin{eqnarray}
\Omega_{\rm DE} = \Omega_\phi+\Omega_{\Lambda} = 1-\Omega_m-\Omega_r
\label{eq:omegade}
\end{eqnarray}

This implies that the \textit{total} dark energy sector, therefore, is dynamical in nature, with an energy density that evolves over cosmic time. Additionally, it is important to realize that only the \textit{total} dark energy density needs to be positive at low redshifts and to satisfy the current observational constraints on the amount of dark energy (i.e., $\Omega_{\rm DE} \approx 0.68$) which is required to account for the late-time accelerated expansion of the Universe \citep{Planck2018}. As a result, the term $\Omega_{\Lambda}$ is, in principle, free to assume any value (positive and negative), thereby allowing for a wide variety of possibilities within the \textit{total} dark energy sector.

From these discussions, it follows that the cosmological parameters - $(w_0, w_a, \Omega_{\Lambda})$ completely characterize the dynamics of the \textit{total} dark energy sector within our cosmological model. In our formalism,  the standard $\Lambda$CDM cosmology corresponds to the case - $w_0 = -1, w_a = 0, \Omega_{\Lambda} =0$ (i.e., $\Omega_\phi = \Omega_{\rm DE}$).

\subsubsection{Abundance of Collapsed Objects}
\label{subsec:abundance_collapsed}

Having specified the background cosmology, we now move towards studying the formation of collapsed objects and their statistical properties at a given cosmic epoch in such a Universe. 

The dark matter halo mass function, which describes the number of dark matter haloes per unit comoving volume at redshift $z$ with masses between $M_h$ and $M_h+dM_h$, can be expressed as follows \citep{Peacock1999_Textbook, Padmanabhan2002_TextbookVol3, Dodelson2020_Textbook}- 
\begin{eqnarray}
\frac{dn}{dM_h}(M_h,z)=-\frac{\bar{\rho}_m}{M_h} \frac{ \mathrm{d}\ln \sigma(M_h,z)}{\mathrm{d}M_h} f\left[\sigma(M_h,z)\right]
\label{eq:dndm}
\end{eqnarray}
where $\bar{\rho}_m$ is the mean comoving background matter density, $\sigma^2(M_h,z)$ is the variance of matter density fluctuations smoothed on the comoving scale $R = \left(3M_h/4\pi \bar{\rho}_m\right)^{1/3}$. The mass variance $\sigma^2(M_h,z)$ is related to the linearly-extrapolated power spectrum of matter density fluctuations $P_{L}(k,z)$ as follows,

\begin{eqnarray}
\sigma^2(M_h,z)=\frac{1}{2\pi^2}\int_0^\infty dk\,k^2 P_L(k,z) \hat{W}^2(k, R)
\label{eq:sigma}
\end{eqnarray}

In the above expression, $\hat{W}(k, R) = 3 [\sin(kR) - kR \cos (kR)] /(kR)^3$ is the Fourier transform of a real-space spherical top-hat window function of radius $R$. The linearly-extrapolated power spectrum $P_L(k,z)$ of matter density fluctuations as a function of wavenumber $k$ at a given redshift $z$ can further be expressed as:
\begin{eqnarray}
P_L(k,z)=P_0 k^{n_s} T^2(k) D^2(z)
\label{eq:pk}
\end{eqnarray}
where $P_0$ is a normalization constant, which is fixed using the present-day rms matter fluctuation amplitude ($\sigma_8$) on a scale of 8 $\hcMpc$, $T(k)$ is the matter transfer function, and $D(z)$ is the linear growth factor. Throughout this work, we normalize the growth factor to $D(z=0)$ = 1. We use the Sheth \& Tormen formalism \citep{ST99} for calculating the halo mass function, wherein the function $f\left[\sigma(M_h,z)\right]$ is given as
\begin{eqnarray}
f\big[\sigma(M_h,z)\big]= A \sqrt{\frac{2a}{\pi}}~\Bigg[ 1+ \left ( \frac{\sigma^2}{a\delta_c^2} \right ) ^p \, \Bigg]~\frac{\delta_c}{\sigma}~\exp \Bigg[ -\frac{a\delta_c^2}{2\sigma^2} \Bigg]
\label{eq:ST_Jenkins}
\end{eqnarray}
with $\delta_c$ = 1.686 representing the critical linear overdensity for collapse. The parameters $(A, a, p)$ are set to the values obtained in  \citet{Jenkins2001}, namely $A=0.353$, $a=0.73$, and $p=0.175$. We use the transfer function, $T(k)$, introduced by  \citet{EisensteinHu1998} in our calculations.

A key ingredient for calculating the halo mass function is the linear growth factor of density perturbations $D(z)$, which determines how density perturbations evolve with redshift in a given cosmological model and is defined in terms of the perturbation amplitude as $ D(z) \equiv \delta(z)/\delta(z=0)$.

We compute the growth factor by numerically solving the second-order differential equation below, which describes the growth of matter density perturbations on sub-Hubble length scales in the linear regime :
\begin{eqnarray}
\delta^{\prime\prime} + \left [ \frac{3}{a}+\frac{E^\prime(a)}{E(a)} \right ]\delta^{\prime} -\frac{3}{2}\frac{\Omega_m}{a^5 E^2(a)}\delta=0\,,
\label{eq:linear_growth}
\end{eqnarray}
where $^\prime$ indicates a derivative with respect to the scale factor $a$ ($\equiv 1/z - 1)$, and $E(a) \equiv H(a)/H_0$ denotes the normalized expansion rate. The last term on the left-hand side of \eqn{eq:linear_growth} represents the gravitational source term, which facilitates the growth of perturbations via the process of gravitational instability, while the second term on the left-hand side of \eqn{eq:linear_growth} corresponds to the Hubble friction term, which tends to suppress their growth due to the expansion of the Universe. The interplay between these competing effects governs the overall rate of structure formation.

Therefore, one can easily see that cosmological models with a dynamical dark energy sector not only alter the background expansion rate (\eqn{eq:hubble_expansion_rate}) but also directly affect the growth of large-scale structures (via modifications to $E(z)$ in \eqn{eq:linear_growth}).   
We solve \eqn{eq:linear_growth} numerically for a given cosmological model, with the initial conditions - $\delta(a_i) = a_i$ and $\delta'(a_i) = 1$ at some initial scale factor $a_i = 10^{-3}$ \citep{Adil2023}.

\begin{figure}
\centering
\includegraphics[width=\columnwidth]{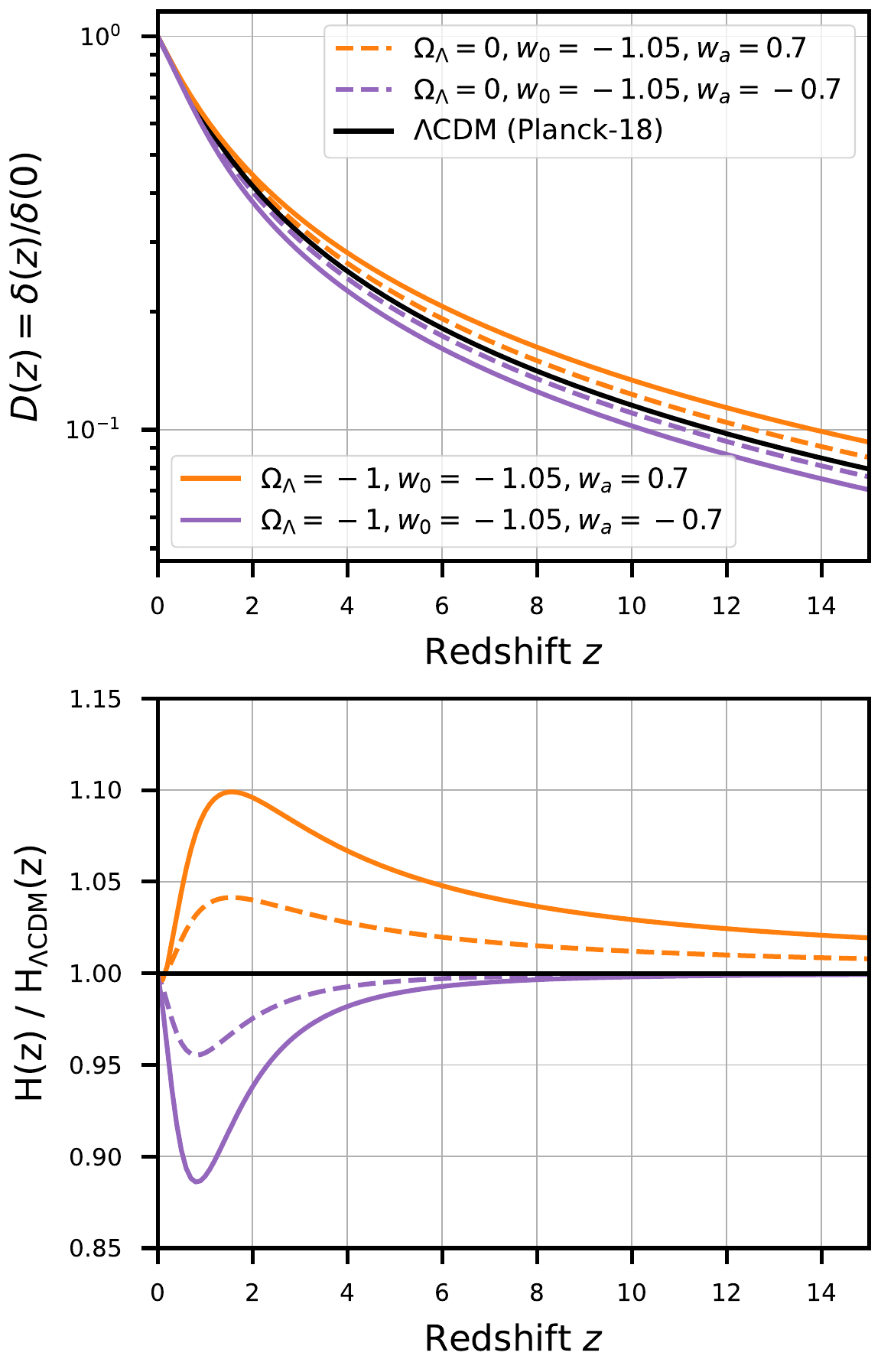}
\caption{The evolution of the linear growth factor (top panel) and the Hubble expansion rate $H(z)$ relative to $\Lambda$CDM (bottom panel) as a function of redshift for different cosmological models with dynamical dark energy.}
\label{fig:Dz_and_Hz_evolution}
\end{figure}

Before proceeding further, let us examine how cosmological models with dynamical dark energy affect the growth of density perturbations and the halo mass function. We only vary the parameters governing the \textit{total} dark energy sector $(w_0, w_a, \Omega_{\Lambda})$, while keeping all other cosmological parameters such as $H_0$, $\Omega_m$, $\Omega_b$, $\Omega_{\rm DE}$, $\sigma_8$, $n_s$ fixed to their \textbf{Planck 2018} best-fit values, as specified at the end of \secn{sec:intro}. For illustration, we select a class of cosmological models, where the evolving CPL component is in the phantom regime at present having $w_\phi < -1 $ (i.e., $w_0 < -1$) but is free to either remain in the phantom regime with $w_\phi < -1 $ (i.e., $w_0 + w_a < -1$) or transition to the non-phantom regime with $w_\phi > -1 $ (i.e., $w_0 + w_a> -1$) at higher redshifts depending on how quickly $w_\phi$ changes with time (i.e., the sign and value of $w_a$).  We show the redshift evolution of the linear growth factor $D(z)$ and the Hubble expansion rate $H(z)$ for such models in the different panels of \fig{fig:Dz_and_Hz_evolution}. In general, for fixed values of $\Omega_{\rm DE}$ and $H_0$, the expansion rate $H(z)$ in these models deviates from that of $\Lambda$CDM due to the evolution of the dark energy density with redshift. At high redshift ($z \gg 1$), for models with $\Omega_\Lambda = 0$ (shown by the dashed curves in the bottom panel of \fig{fig:Dz_and_Hz_evolution}), the expansion rate exceeds that of $\Lambda$CDM once the evolving energy density of the field $\phi$ becomes larger than that of a cosmological constant. This condition can be quantified by the ratio $R(z) \equiv \rho_\phi/\rho_\Lambda
\simeq (1+z)^{3(1+w_0+w_a)}\,e^{-3w_a}$, such that $H(z) > H_{\Lambda\mathrm{CDM}}(z)$ when $R(z) > 1$. It follows from \eqn{eq:linear_growth} that a faster expansion rate (compared to the $\Lambda$CDM model) suppresses the growth of matter perturbations due to increased Hubble friction. When normalized to produce the same amplitude today, these fluctuations must grow more rapidly at early times to reach this value by $z=0$, resulting in larger growth factors at higher redshifts, as shown in the top panel of \fig{fig:Dz_and_Hz_evolution}. By similar arguments, models that exhibit an expansion rate slower than $\Lambda$CDM yield smaller growth factors. The inclusion of a negative cosmological constant only aggravates the impact on the growth factor at high redshifts since a more negative $\Lambda$ (i.e., a larger value of $|\Omega_\Lambda|$) demands a correspondingly larger positive energy density contribution from the scalar field component ($\Omega_\phi$) to satisfy observational constraints on late-time acceleration—namely, that $\Omega_\Lambda + \Omega_\phi \simeq 0.68$ (see \eqn{eq:omegade}).

\begin{figure*}	
\centering
\includegraphics[width=\textwidth]{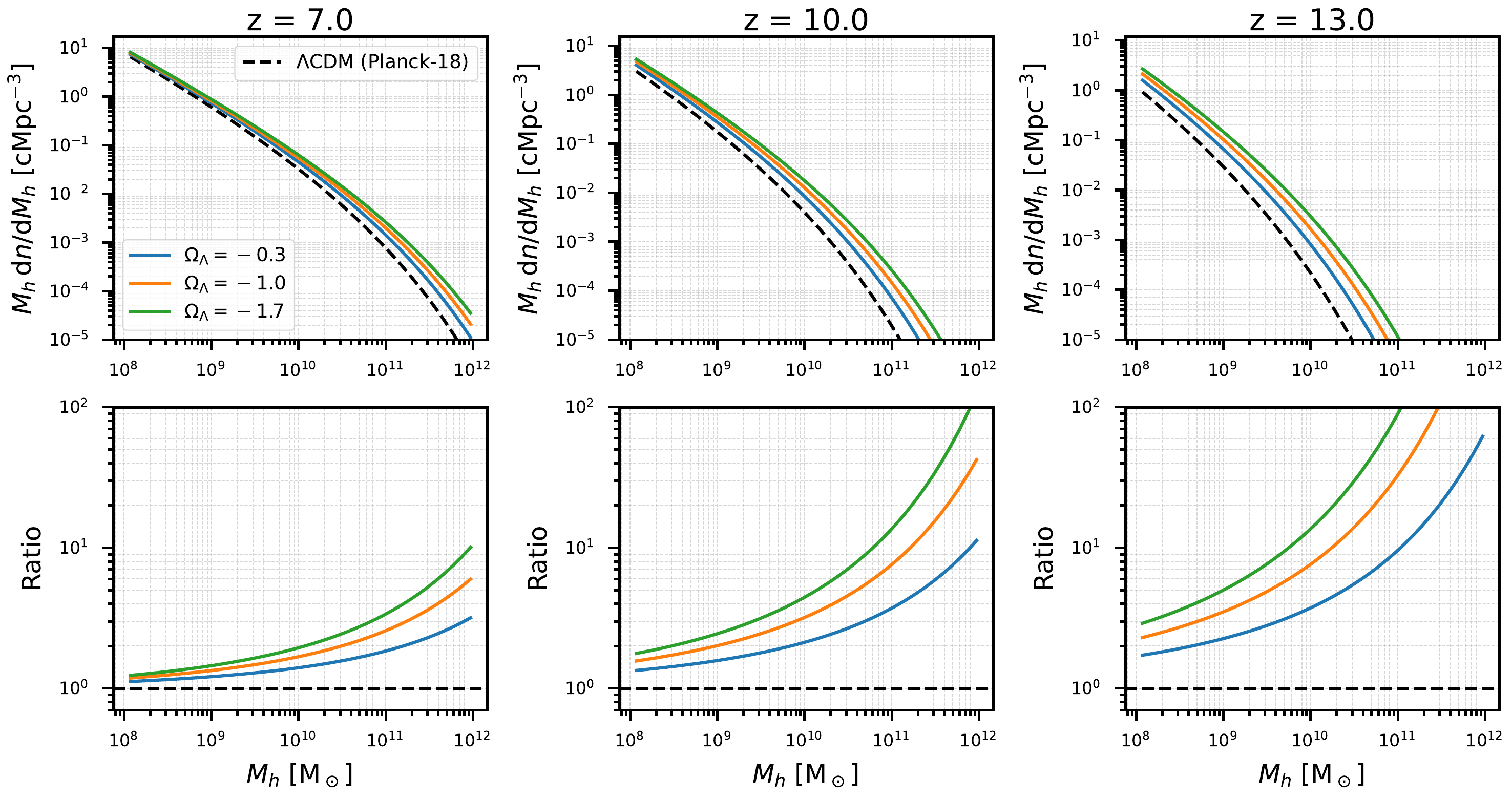}	
\caption{The effect of dynamical dark energy models, featuring a scalar field with a redshift-dependent equation of state described by \eqn{eq:cpl_eos} and a negative cosmological constant $\Lambda$, on the dark matter halo mass functions at high redshifts ($z=7, 10,$ and $13$). Here, we only vary $\Omega_\Lambda$ while keeping the equation of state parameters of the scalar field $\phi$ fixed ($w_0 = -1.05$, $w_a = 0.7$).}
\label{fig:HMF_for_nCC}
\end{figure*}

In \fig{fig:HMF_for_nCC}, we show the impact of increasingly negative values of the cosmological constant in the dark energy sector (alongside an evolving scalar field component having fiducial values of $w_0 = -1.05$ and $w_a = 0.7$) on the halo mass function at $z > 6$. As previously discussed, a more negative cosmological constant necessitates a proportionally higher energy density in the evolving scalar field component ($\Omega_\phi$), which in turn further enhances the growth factor at earlier times, ultimately leading to a pronounced boost in the halo mass function. In the cosmological models chosen for illustration in \fig{fig:HMF_for_nCC}, we find that the abundance of dark matter haloes of a fixed mass compared to the $\Lambda$CDM model systematically increases as one moves towards higher redshifts.

Having outlined the underlying cosmological framework, we now turn to describing the astrophysical modelling of UV-bright galaxies that form within these dark matter haloes in the next subsection.

\subsection{Astrophysical Model: Populating Haloes with UV Galaxies}
\label{subsec:galform_framework}

In this subsection, we present the astrophysical framework of our theoretical model, outlining the prescriptions used to model star formation activity and ionizing photon production in high-redshift galaxies, as well as the methodology used to compute various global observables associated with galaxy populations and cosmic reionization. We use the framework described in our earlier works \citep{Chakraborty2024, Chakraborty2026} (hereafter, \citetalias{Chakraborty2024} and \citetalias{Chakraborty2026}), which self-consistently models the evolving UV luminosity function (UVLF) of galaxies and the global reionization history while incorporating the effects of radiative feedback. We present a brief summary of the main features of the model below, and refer interested readers to \citetalias{Chakraborty2024} and \citetalias{Chakraborty2026} for further details.

In this model, the star-formation rate $\dot{M}_*$ of a galaxy residing within a halo of mass $M_h$ is calculated as
\begin{eqnarray}
\label{eq:SFR}
\dot{M}_*(M_h,z) = \dfrac{f_*(M_h,z)}{t_\ast(z)}~f_{\rm gas}(M_h)~ \bigg(\dfrac{\Omega_b}{\Omega_m}\bigg) M_h ,
\end{eqnarray}
where $f_*(M_h,z)$ represents the star-formation efficiency (i.e., the fraction of baryons within haloes that are converted into stars), $f_{\rm gas}(M_h)$ denotes the gas fraction retained inside a halo after photo-heating/photo-evaporation due to the rising ionizing UV background, and $t_\ast(z) = c_\ast~t_H(z)$ is the average star formation time scale, with $t_H(z) = 1/H(z)$ being the local Hubble time and $c_\ast$ a dimensionless constant. If star formation occurs on a timescale comparable to the halo free-fall time ($t_\ast \simeq t_{\rm ff}$), this implies $c_\ast \approx 0.11$ for a virial overdensity $\Delta_c \approx 18\pi^2$ \footnote{The free-fall time of a virialized dark-matter halo, $t_{\rm ff} = t_{\rm dyn}/\sqrt{2} = \sqrt{3\pi/(32G\bar{\rho}_{\rm vir})}$, can be written in terms of the Hubble time as $t_{\rm ff}(z) = \pi/(2\sqrt{\Delta_c})\,t_H(z)$, assuming $\bar{\rho}_{\rm vir}(z) = \Delta_c\,\rho_c(z)$. For the virial overdensity $\Delta_c \approx 18\pi^2$, this yeilds $t_{\rm ff}(z)\approx 0.11\,t_H(z)$.}.
For haloes that form within already ionized regions and are thus subject to radiative feedback, the gas fraction is modeled as $f_{\rm gas}(M_h) = 2^{-\Mcrit/M_h}$, where $\Mcrit$ denotes the characteristic halo mass capable of retaining 50\% of its baryonic gas reservoir \citep{Sobacchi&Mesinger_1Dsims_2013}. In contrast, for haloes located in neutral regions where radiative feedback is absent, the gas fraction is assumed to be unity, i.e., $f_{\rm gas}(M_h) = 1$.

The monochromatic rest-frame UV luminosity ($\LUV$) of a galaxy at 1500 \angstrom, which is derived from its star formation rate\footnote{The star formation rate and UV luminosity are related through the relation $\LUV = \dot{M}_*(M_h,z) / \kappa_{\rm UV}$, wherein the constant conversion factor $\kappa_{\rm UV}$ depends on the star formation history and characteristics of the stellar population, such as its age, metallicity, binarity, and the initial mass function (IMF).}, is likewise also modulated by the effects of radiative feedback during reionization. As a result, the relationship between halo mass and UV luminosity depends on whether a galaxy resides in an ionized or neutral region of the intergalactic medium.
For galaxies that are unaffected by radiative feedback, their UV luminosity is determined as follows,
\begin{eqnarray}
\label{eq:LUV_nofb}
L^{\rm nofb}_{{\rm UV}}(M_h,z) = \dfrac{\dot{M}^{\rm nofb}_*(M_h,z)}{\mathcal {K}_{{\rm UV}}} = \dfrac{\varepsilon_{{\rm *,UV}}(M_h,z)}{\mathcal {K}_{{\rm UV,fid}}}~\bigg(\dfrac{\Omega_b}{\Omega_m}\bigg) M_h
\end{eqnarray} 

Whereas in regions that have already been ionized, the associated UV background suppresses star formation in low-mass haloes. The UV luminosity of galaxies in these feedback-affected regions is calculated as:
\begin{align}
\label{eq:LUV_fb}
L^{\rm fb}_{{\rm UV}}(M_h,z) 
&= 2^{-\Mcrit/M_h}~L^{\rm nofb}_{{\rm UV}}(M_h,z) \nonumber\\
&= 2^{-\Mcrit/M_h}~\dfrac{\varepsilon_{{\rm *,UV}}(M_h,z)}{\mathcal {K}_{{\rm UV,fid}}}~\bigg(\dfrac{\Omega_b}{\Omega_m}\bigg) M_h
\end{align}

In these equations, the parameter $\varepsilon_{{\rm *,UV}}(M_h,z)$ denotes the UV efficiency of the halo and encapsulates several key parameters discussed above that govern star formation and the resulting ultraviolet emission in galaxies: 
\begin{eqnarray}
\label{eq:epsilonstarUV}
\varepsilon_{{\rm *,UV}}(M_h,z) \equiv \dfrac{f_\ast(M_h,z)}{c_\ast~t_H({z)}} \, \dfrac{\mathcal{K}_{\rm UV,fid}}{\mathcal{K}_{\rm UV}}
\end{eqnarray}

Assuming a star-formation efficiency parameterized as $f_{\ast}(M_h)= f_{*,10} \big(M_h/10^{10} M_\odot \big)^{\alpha_*}$, the UV efficiency parameter introduced in \eqn{eq:epsilonstarUV} takes the form -
\begin{eqnarray}
\label{eq:epsilonstarUV_simplified}
\varepsilon_{{\rm *,UV}}(M_h,z)  = \dfrac{\varepsilon_{{\rm *10,UV}}}{t_H(z)} \left(\dfrac{M_h}{10^{10} M_\odot}\right)^{\alpha_\ast} ,
\end{eqnarray}
where the normalization $\varepsilon_{{\rm *10,UV}}$ is defined as : 
\begin{eqnarray}
\label{eq:epsilonUVstar_norm}
\varepsilon_{{\rm *10,UV}} \equiv \dfrac{f_{*,10}}{c_{*}}~\dfrac{\mathcal {K}_{\rm UV,fid}}{\mathcal {K}_{\rm UV}}.
\end{eqnarray}

The rest-frame UV luminosities obtained from this model are finally converted to absolute UV magnitudes (in the AB system) using the relation \citep{Oke_ABmag, Oke&Gunn_ABmag} - 
\begin{eqnarray}
{\rm log_{10}}\left(\frac{L_{\rm UV}}{{\rm erg \ s^{-1} \ Hz^{-1}}} \right) = 0.4 \times (51.6 - M_{\rm UV}). 
\end{eqnarray} 

Given a halo mass function and assuming each dark matter halo hosts only one galaxy, it is relatively straightforward to calculate the UV luminosity function using these $M_h-\LUV$ relations mentioned in \eqns{eq:LUV_nofb}{eq:LUV_fb}. The globally averaged UV luminosity function (${\rm \Phi^{total}_{UV}}$) at a given redshift $z$ is then obtained by appropriately combining the feedback-affected UV luminosity function (${\rm \Phi^{fb}_{UV}}$) from ionized regions and the feedback-unaffected UV luminosity function (${\rm \Phi^{nofb}_{UV}}$) from neutral regions, as follows - 

\begin{align}
\label{eqn:lumfunc_full}
\Phi^{\rm total}_{\rm UV}(\MUV, z) 
&= Q_{\rm HII}(z)\, \Phi^{\rm fb}_{\rm UV} 
   + \big[1 - Q_{\rm HII}(z)\big]\, \Phi^{\rm nofb}_{\rm UV} \notag\\
&= Q_{\rm HII}(z)\, \frac{{\rm d}n}{{\rm d}M_h} 
   \left|\frac{{\rm d}M_h}{{\rm d}L^{\rm fb}_{\rm UV}}\right|
   \left|\frac{{\rm d}L^{\rm fb}_{\rm UV}}{{\rm d}M_{\rm UV}}\right| \notag\\
&\quad ~~~ + \big[1 - Q_{\rm HII}(z)\big]\,
   \frac{{\rm d}n}{{\rm d}M_h}
   \left|\frac{{\rm d}M_h}{{\rm d}L^{\rm nofb}_{\rm UV}}\right|
   \left|\frac{{\rm d}L^{\rm nofb}_{\rm UV}}{{\rm d}M_{\rm UV}}\right|
\end{align}
where  $Q_{\rm HII}(z)$ is the globally averaged ionization fraction at redshift $z$ and ${\rm d}n / {\rm d}M_h$ is the dark matter halo mass function. 

As evident from \eqn{eqn:lumfunc_full}, the globally averaged ionized hydrogen fraction, $Q_{\rm HII}$, constitutes a critical input for determining the UV luminosity function. In our model, we self-consistently follow the evolution of $Q_{\rm HII}$, whose rate of change is controlled by the net balance between ionization and recombination processes in the intergalactic medium.  For this calculation, the intrinsic ionizing photon production rate within a dark matter halo is modeled in terms of its star formation rate and the number of ionizing photons emitted per unit stellar mass formed ($\etaGamma$). However, not all ionizing photons produced within a halo can leak out and reach the intergalactic medium. We assume that the fraction of hydrogen ionizing photons that escapes into the intergalactic medium depends on the halo mass and is parameterized as $f_{\rm esc}(M_h)= f_{\rm esc,10} \big(M_h/10^{10} M_\odot \big)^{\alpha_{\rm esc}}$. 

Hence, the number density of ionizing photons per unit time contributed by feedback-unaffected galaxies is given by 
\begin{multline}
    \dot{n}^{\rm nofb}_{\rm ion} (z) =\eta_{\gamma*,{\rm fid}}  \int_{M_{\rm cool}(z)}^{\infty} \varepsilon_{{\rm esc}}(M'_h,z)~~\varepsilon_{{\rm *,UV}}(M'_h,z) ~\left(\frac{\Omega_b}{\Omega_m}\right) ~~ \\ \times M'_h ~~ \frac{\mathrm{d}n}{\mathrm{d}M_h}(M'_h,z) ~~ \mathrm{d}M'_h
\label{eq:niondot_nofb}
\end{multline}

In contrast, the contribution from galaxies residing within ionized regions is regulated by radiative feedback and is computed as
\begin{multline}
\dot{n}^{\rm fb}_{\rm ion} (z) = \eta_{\gamma*,{\rm fid}} \int_{M_{\rm cool}(z)}^{\infty} ~2^{-\Mcrit/M'_h}~~\varepsilon_{{\rm esc}}(M'_h,z)~~\varepsilon_{{\rm *,UV}}(M'_h,z) ~ \\ \times \left(\frac{\Omega_b}{\Omega_m}\right) ~~ M'_h ~~ \frac{\mathrm{d}n}{\mathrm{d}M_h}(M'_h,z) ~~ \mathrm{d}M'_h
\label{eq:niondot_fb}
\end{multline}

Here, $M_{\rm cool}(z)$ denotes the halo mass corresponding to the atomic cooling threshold (i.e., $T_{\rm vir} = 10^4$ K) at redshift $z$ and $\varepsilon_{{\rm esc}}(M_h,z)$ represents the escaping ionizing efficiency, defined as :
\begin{align}
\label{eqn:espilon_escape}
\varepsilon_{{\rm esc}}(M_h,z) &\equiv \dfrac{\kappaUV}{\kappaUVfid} ~ \dfrac{\etaGamma}{\etaGammafid}~f_{{\rm esc}}(M_h,z)  \nonumber \\
&= \dfrac{\xiIon}{\xiIonfid}~f_{{\rm esc}}(M_h,z).
\end{align}

Given our assumption of $f_{\rm esc}(M_h)$ being a power-law function, the escaping ionizing efficiency can be further expressed as  
\begin{eqnarray}
\label{eqn:espilon_escape_simplified}
\varepsilon_{\rm esc}(M_h,z) = \varepsilon_{{\rm esc,10}} \left(\frac{M_h}{10^{10} M_\odot}\right)^{\alpha_{\rm esc}},
\end{eqnarray}
where   
\begin{eqnarray}
\label{eqn:espilon_escape_normalisation}
\varepsilon_{{\rm esc,10}} \equiv \dfrac{\xiIon}{\xiIonfid}  f_{{\rm esc,10}}.
\end{eqnarray}

The \emph{total} comoving number density of ionizing photons that escapes into the intergalactic medium per unit time, and sources the growth of ionized regions, is therefore given by
\begin{eqnarray}
\label{eq:niondot_total}
\dot{n}_{\rm ion}(z) = Q_{\rm HII}(z)~\dot{n}^{\rm fb}_{\rm ion}(z) + [1- Q_{\rm HII}(z)]~\dot{n}^{\rm nofb}_{\rm ion}(z) 
\end{eqnarray}

We adopt a fiducial value of $\mathcal{K}_{{\rm UV,fid}} = 1.15485 \times 10^{-28}~{\rm \mathrm{M}_\odot}\ {\rm yr}^{-1}/ ({\rm erg}~{\rm s}^{-1}~{\rm Hz}^{-1})$ and $\eta_{\gamma*,{\rm fid}}$ = $4.62175 \times 10^{60}$ photons per M$_\odot$ in all our calculations. These values were obtained using {\tt{STARBURST99 v7.0.1}}\footnote{\url{https://www.stsci.edu/science/starburst99/docs/default.htm}}\citep{Starburst99} for a stellar population with a Salpeter IMF  (0.1 - 100 $\Msun$) and metallicity $Z = 0.001 ( = 0.05~Z_\odot)$ at an age of 100 Myr, assuming continuous star formation. The assumed fiducial values for $\mathcal {K}_{{\rm UV}}$ and $\eta_{\gamma*}$ correspond to an ionizing photon production efficiency $\log_{10} \big[\xi_{\rm ion,fid}/({\rm erg}^{-1}\ {\rm Hz}) \big] \approx 25.23$, which is consistent with the latest measurements from JWST \citep{Simmonds2024_JADES_xiION, Pahl2025_xiION, Begley2025_xiION}.

Once the global reionization history $Q_{\rm HII}(z)$ is obtained, the Thomson scattering optical depth of the CMB photons for that particular model can be computed as
\begin{eqnarray}
\label{eq:tauCMB}
\tau_{\rm el} = \sigma _T \bar{n}_{H}c \int _0^{z_{\rm LSS}} \frac{\mathrm{d}z^{\prime }}{H(z^{\prime })} ~ (1 + z^{\prime })^2 ~ \chi _{\mathrm{He}}(z^{\prime }) ~ Q_{\mathrm{HII}}(z^{\prime }),
\end{eqnarray}
where $z_{\rm LSS}$ is the redshift of last scattering,  $\bar{n}_{\rm H} $ is the current mean comoving number density of hydrogen, and $\sigma_T$ is the Thomson scattering cross-section. 

Before proceeding ahead, we note that our theoretical model includes \textbf{five} free parameters that describe the astrophysical properties of high-redshift galaxies --- namely, \( \log_{10}(\varepsilon_{\ast10,\mathrm{UV}}) \), \( \alpha_\ast \), \( \log_{10}(\varepsilon_{\mathrm{esc},10}) \), \( \alpha_{\mathrm{esc}} \), and  \( \log_{10}(M_{\mathrm{crit}}/M_\odot) \).

\section{Observational Datasets and Likelihood Analysis}
\label{sec:data_and_methods}

We compare the theoretical predictions of the model described in the previous section with several available observational datasets. In this section, we briefly summarize them and also describe the Bayesian formalism used to constrain the free parameters of our model.

\begin{enumerate}
    \item \textbf{Thomson scattering optical depth of CMB photons: } Throughout this work (except in \secn{subsec:results_maxboost_nCC_model}), we use the latest measurement of $\tau_\mathrm{el}= 0.0540 \pm 0.0074$ reported by the Planck collaboration \citep{Planck2018}.

    \item  \textbf{Global Reionization History: } We utilize estimates of the globally averaged fraction of intergalactic neutral hydrogen ($Q_\mathrm{HI} = 1 - Q_\mathrm{HII} $) at different redshifts derived from analyses of Lyman-$\alpha$ absorption features in the spectra of distant quasars and galaxies using hydrodynamical or semi-numerical simulations, similar to our previous work \citepalias{Chakraborty2024,Chakraborty2026}. At $z \lesssim 6$, constraints on $Q_{\rm HI}$ are obtained by comparing the Lyman-$\alpha$ forest opacity fluctuations predicted by the simulations with those measured in quasar spectra \citep{Gaikwad2023}. At higher redshifts ($z > 6$), where the Lyman-$\alpha$ forest becomes saturated, the constraints on $Q_{\rm HI}$ are instead determined by modelling the Lyman-$\alpha$ damping-wing absorption observed in the spectra of high-redshift galaxies and quasars \citep{Davies2018, Greig2022, Umeda2023, Durovcikova2024}.
   
    \item \textbf{Galaxy UV Luminosity Functions:}  We make use of observational data for the galaxy UV luminosity function, $\Phi_{\rm UV} (\MUV, z)$, compiled across nine redshifts in the range $5 \leq z < 14$. These measurements are obtained from a combination of surveys carried out with the Hubble Space Telescope \citep{Bouwens2021} and the James Webb Space Telescope \citep{Donnan2023, Harikane2023, Bouwens2023, McLeod2024, Donnan2024}.

    Following our previous work \citepalias{Chakraborty2024,Chakraborty2026}, we only consider the observational data points with $M_{\rm UV} \geq -21$ from these studies in our analysis since our theoretical model does not incorporate the effects of feedback from active galactic nuclei (AGN) activity or the significant dust attenuation present in bright galaxies \citep{Mauerhofer&Dayal2023}.

\end{enumerate}

We adopt a Bayesian framework to constrain the free parameters of our model by comparing its predictions against the complete set of observational constraints discussed above. In this approach, we compute the posterior probability distribution, $\mathcal{P}(\boldsymbol{\theta} \vert \mathcal{D})$, of the model parameters $\boldsymbol{\theta}$ given the data $\mathcal{D}$, using Bayes’ theorem:
\begin{eqnarray}
\mathcal{P}(\boldsymbol{\theta} \vert \mathcal{D}) = \frac{\mathcal{L}(\mathcal{D} \vert \boldsymbol{\theta})~ \pi(\boldsymbol{\theta})}{\mathcal{Z}},
\end{eqnarray}
where $\mathcal{L}(\mathcal{D} \vert \boldsymbol{\theta})$ is the likelihood function, representing the conditional probability distribution of the data $\mathcal {D}$ given the model parameters $\boldsymbol \theta$; $\pi(\boldsymbol{\theta})$ is the prior distribution of the model parameters; and $\mathcal {Z} = \int \mathcal {L}(\mathcal {D} \vert \boldsymbol \theta) ~\pi (\boldsymbol \theta) ~ d \boldsymbol \theta$ is the Bayesian evidence. Since our focus is on parameter estimation rather than model comparison, the evidence $\mathcal{Z}$ serves only as a normalization constant and plays no role in our analysis.

Assuming the different observational datasets to be statistically independent, the joint total likelihood is calculated as the product of the individual likelihoods:
\begin{eqnarray}
\label{eqn:joint_likelihood_expression}
\mathcal {L}(\mathcal {D} \vert \boldsymbol \theta) =  \prod_\alpha \mathcal {L}(\mathcal {D}_\alpha \vert \boldsymbol \theta),
\end{eqnarray}
where $\alpha$ indexes the individual datasets included in the analysis. For each dataset $\mathcal{D}_\alpha$, the likelihood is assumed to be Gaussian, and can therefore be written as
\begin{align}
\label{eqn:individual_likelihood_expression}
\mathcal{L}(\mathcal{D}_\alpha \vert \boldsymbol{\theta}) 
&= \exp\left[-\frac{1}{2}~\chi^2(\mathcal{D}_\alpha, \boldsymbol{\theta})\right] \nonumber \\
&= \exp\left[-\frac{1}{2} \sum_i \left( \frac{\mathcal{D}_{\alpha,i} - \mathcal{M}_{\alpha,i}(\boldsymbol{\theta})}{\sigma_{\alpha,i}} \right)^2 \right],
\end{align}
where $\mathcal{D}_{\alpha,i}$ and $\sigma_{\alpha,i}$ are the observed value and its associated uncertainty for the $i$-th data point, respectively, and $\mathcal{M}_{\alpha,i}(\boldsymbol{\theta})$ is the model prediction corresponding to that data point. In cases where the data is reported with asymmetric errorbars, we evaluate the likelihood using the upper (lower) uncertainty when the model prediction lies above (below) the observed value.

In order to facilitate a fair comparison between the predictions from our CPL$n\Lambda$CDM models and the observed galaxy UV luminosities and number densities reported in these studies, originally derived from the flux and number counts respectively under the assumption of a specific cosmological model (viz., $\Lambda$CDM), we need to make relevant corrections that would convert it to what they would be if interpreted by an observer assuming a $\Lambda$CDM cosmology. The ``corrected''  UV luminosity function $\Phi_{\rm UV}^{\prime}$ and UV magnitude $M_{\rm UV}^{\prime}$ are given by,

\begin{align}
    \Phi_{\rm UV}^{\prime} &= \Phi_{\rm UV} \times \frac{({\rm d}V/{\rm d}z)_{\mathrm{CPL}n\Lambda \mathrm{CDM}
}}{({\rm d}V/{\rm d}z)_{\Lambda{\rm CDM}}}, \\
    M_{\rm UV}^{\prime} &= M_{\rm UV} - 2.5 \log_{10} \left[ \left( \frac{D^{\Lambda{\rm CDM}}_{\rm L}}{D^{\mathrm{CPL}n\Lambda \mathrm{CDM}}_{\rm L}} \right)^2 \right],
\end{align}

where ${\rm d}V/{\rm d}z$ and $D_{\rm L}$ are the differential comoving volume and luminosity distance at the redshift of interest.

Throughout this paper, all galaxy number densities and UV luminosities shown in the figures and discussed in the text for the CPL$n\Lambda$CDM model will refer to these corrected quantities.

Another important caveat in our analysis involves the treatment of the neutral hydrogen fraction ($Q_{\rm HI}$) measurements obtained by various observational studies. We assume that these constraints are primarily sensitive to the astrophysical modelling rather than the underlying cosmological model (viz., $\Lambda$CDM) assumed in these works. Since reinterpreting these $Q_{\rm HI}$ measurements within a different cosmological framework is non-trivial, we adopt the published values without modification in our likelihood calculations for the CPL$n\Lambda$CDM models to keep the analysis simple.

\section{Results and Discussion}
\label{sec:results}


\subsection{A fiducial model: success with JWST, tension with the CMB}
\label{subsec:results_fiducial_nCC_model}

\begin{figure*}
\centering
\includegraphics[width=\textwidth]{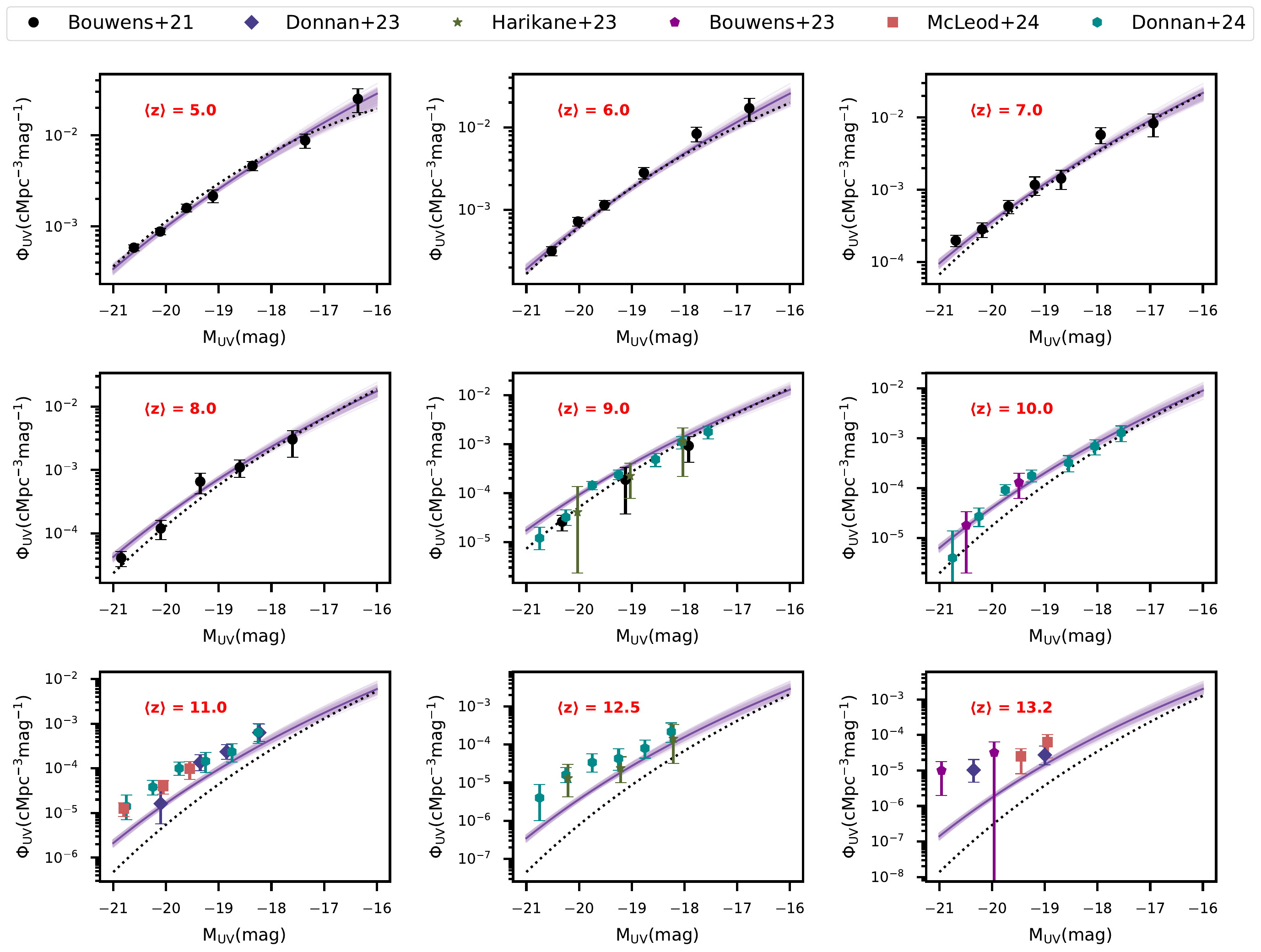}
\caption{The galaxy UV luminosity functions at nine different redshifts (with their respective mean values $\langle z \rangle$ mentioned in the upper left corner) for 200 random samples drawn from the MCMC chains of the \textbf{fiducial CPL$\bm{n\Lambda}$CDM} case. In each panel, the solid dark-violet line corresponds to the best-fitting model, while the colored data points show the different observational constraints \citep{Bouwens2021, Donnan2023, Harikane2023, Bouwens2023, McLeod2024, Donnan2024} used in the likelihood analysis. The prediction from a model within the \textbf{$\bm{\Lambda}$CDM (Planck 2018)} cosmology that best matches the observational measurements at $z < 10$ and does not assume any evolution in the UV efficiency parameters above $z \sim 10$ is also shown using black dotted lines.}
\label{fig:fiducial_UVLF}
\end{figure*}

\begin{figure}
\centering
\includegraphics[width=\columnwidth]{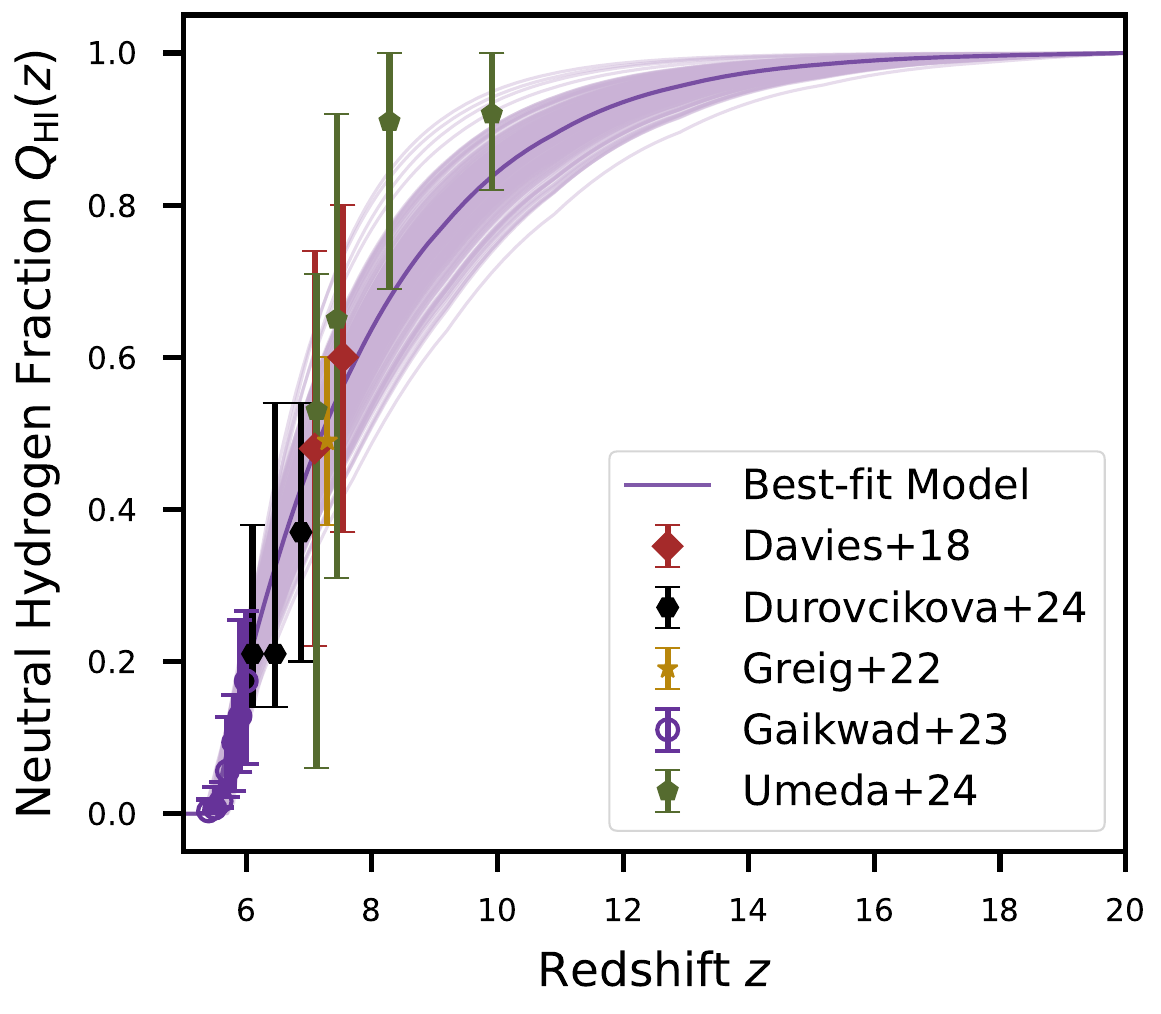}
\caption{The evolution of the globally averaged intergalactic neutral hydrogen fraction as a function of redshift for 200 random samples drawn from the MCMC chains of the \textbf{fiducial CPL$\bm{n\Lambda}$CDM} case. The colored data points represent the observational measurements of $Q_{\rm HI}$($z$) used in the analysis. } 
\label{fig:fiducial_reion_history}
\end{figure}

\begin{table*}
 \caption{Constraints on the \textit{astrophysical} free parameters in the CPL$n\Lambda$CDM cosmological model from the MCMC-based analysis. The parameters are assumed to follow uniform priors within the ranges specified in the second column. The numbers in the other columns show the mean value with 1$\sigma$ errors for these parameters, as obtained for the two cases described in \secn{subsec:results_fiducial_nCC_model} and \secn{subsec:results_maxboost_nCC_model} respectively. Note that $\tau_{\rm el}$ is a derived parameter in our analysis.\\} 
 \label{tab:mcmc_results}
\begin{tabular*}{0.75\textwidth}{c@{\hspace*{30pt}}c@{\hspace*{30pt}}c@{\hspace*{30pt}}c@{\hspace*{30pt}}}
\hline \hline \\
Parameters & Priors &
\parbox[c]{3.5cm}{\centering fiducial\\\hspace*{1em}CPL$n\Lambda$CDM} &
\parbox[c]{3.5cm}{\centering maxboost\\\hspace*{1em}CPL$n\Lambda$CDM} \\ \\
\hline \hline \\

{\boldmath$\log_{10}~(\varepsilon_{\mathrm{\ast 10, UV}})$}& [-3.0, 0.5] & $-1.224\pm 0.033$&  $-0.813^{+0.032}_{-0.028}$ \\ \\

{\boldmath$\alpha_\ast$} & [-1.0, 1.0]  & $0.338\pm 0.031$ & $0.228^{+0.026}_{-0.029}$   \\ \\

{\boldmath$\log_{10}~(\varepsilon_{\mathrm{esc,10}})$} & [-3.0, 1.0]  & $-0.822\pm 0.047$ & $-0.841\pm 0.028$ \\ \\

{\boldmath$\alpha_{\rm esc}$} & [-3.0, 1.0] & $-0.23\pm 0.12$ & $-0.087^{+0.063}_{-0.072}$  \\ \\

{\boldmath$\log_{10}(M_{\mathrm{crit}}/M_\odot)$} & [9.0, 11.0] & $< 9.62$ & $10.496^{+0.096}_{-0.081}$ \\ \\

\hline \\
$\tau_{\rm el} $ & - &  $0.0548^{+0.0029}_{-0.0032}$ & $0.0597^{+0.0028}_{-0.0031}$ \\ \\

\hline
\end{tabular*}
\end{table*}

In our earlier works \citepalias{Chakraborty2024,Chakraborty2026}, we explained the various high-redshift observations discussed in \secn{sec:data_and_methods} by invoking a redshift evolution in the astrophysical parameters governing galaxy formation and evolution, while assuming a $\Lambda$CDM background cosmology. For completeness, we summarize the constraints obtained from such an analysis in \app{appendix:LCDM_astro_evolution} using the \textbf{$\bm{\Lambda}$CDM (Planck 2018)} cosmological parameters.

However, in this work, our aim is to assess the viability of cosmological models beyond $\Lambda$CDM in explaining these high-redshift datasets, \textit{without invoking any evolution in the astrophysical properties of early galaxies}. For this purpose, we choose a fiducial model in the CPL$n\Lambda$CDM cosmology, whose dark energy sector is characterized by the parameters - $\Omega_\Lambda = -1, w_0 = -1.05, w_a = 0.7$.  As has been the recent practice in literature \citep{Menci2020, Adil2023, Menci2024, Wang2024_DDE}, we fix all other relevant cosmological parameters — such as $H_0$, $\Omega_m$, $\sigma_8$, and $n_s$ — to their \textbf{$\bm{\Lambda}$CDM (Planck 2018)} best-fit values described at the end of \secn{sec:intro}, while noting that this approach, in principle, ignores potential correlations between cosmological parameters.
From \figs{fig:Dz_and_Hz_evolution}{fig:HMF_for_nCC}, we have already seen that the number density of dark matter haloes at $z>10$ in this  fiducial CPL$n\Lambda$CDM cosmology is significantly larger than in the $\Lambda$CDM cosmology.

We therefore perform a  Markov chain Monte Carlo (MCMC) analysis to constrain the \textit{astrophysical} parameters for this cosmological model by comparing its theoretical predictions with the various observations listed in the previous section. We shall henceforth refer to this case as the \textbf{``~fiducial CPL$\bm{n\Lambda}$CDM~''} model. The marginalized constraints on the free parameters are mentioned in the third column of \tab{tab:mcmc_results}. We show the model-predicted UVLFs for 200 random samples from the MCMC chains in \fig{fig:fiducial_UVLF}, and their corresponding reionization histories in \fig{fig:fiducial_reion_history}.

A natural consequence of a larger abundance of haloes at higher redshifts in the \textbf{fiducial CPL$\bm{n\Lambda}$CDM} cosmological model compared to \textbf{$\bm{\Lambda}$CDM (Planck 2018)} is an increased number density of massive UV-luminous galaxies at high redshifts, particularly at $z>10$.  As a result, this model shows noticeably better agreement with the UVLF observations at $z>10$ than the corresponding predictions from the $\Lambda$CDM-based model, for which a redshift-independent UV production efficiency is assumed (shown as black dotted lines in \fig{fig:fiducial_UVLF}). We find that reproducing the observed UVLFs over the redshift range $5 \leq z < 14$ in this cosmological framework requires a relatively lower UV efficiency in galaxies on average ($\varepsilon_{\ast10,\mathrm{UV}} \approx 0.059$) compared to that required in $\Lambda$CDM cosmology, where $\varepsilon_{\ast10,\mathrm{UV}} \approx 0.13$ at $z \lesssim 9$ and increases further at higher redshifts (see \app{appendix:LCDM_astro_evolution}). However, as shown in \fig{fig:fiducial_UVLF}, the \textbf{fiducial CPL$\bm{n\Lambda}$CDM} model still falls short in reproducing the full shape of the observed UVLFs at $z \geq 11$, particularly at the bright end. This suggests that introducing a modest redshift evolution in the astrophysical parameters -- potentially, in the slope of the halo mass–stellar mass relation ($\alpha_\ast$) -- may be essential for bringing the UVLF predictions of the \textbf{fiducial CPL$\bm{n\Lambda}$CDM} cosmological model into better agreement with the observations.

As seen in \fig{fig:fiducial_reion_history}, the \textbf{fiducial CPL$\bm{n\Lambda}$CDM} model also successfully reproduces the various observational constraints on the progress of reionization. We obtain an escaping ionizing efficiency of $\approx 15\%$ for $10^{10} M_\odot$ haloes and find $\varepsilon_{\rm esc}$ to be negatively correlated with halo mass, similar to the trends obtained for reionization-era galaxies within the $\Lambda$CDM cosmological model (\tab{tab:LCDM_astro_evolve_results}).

\begin{figure}
\centering
\includegraphics[width=\columnwidth]{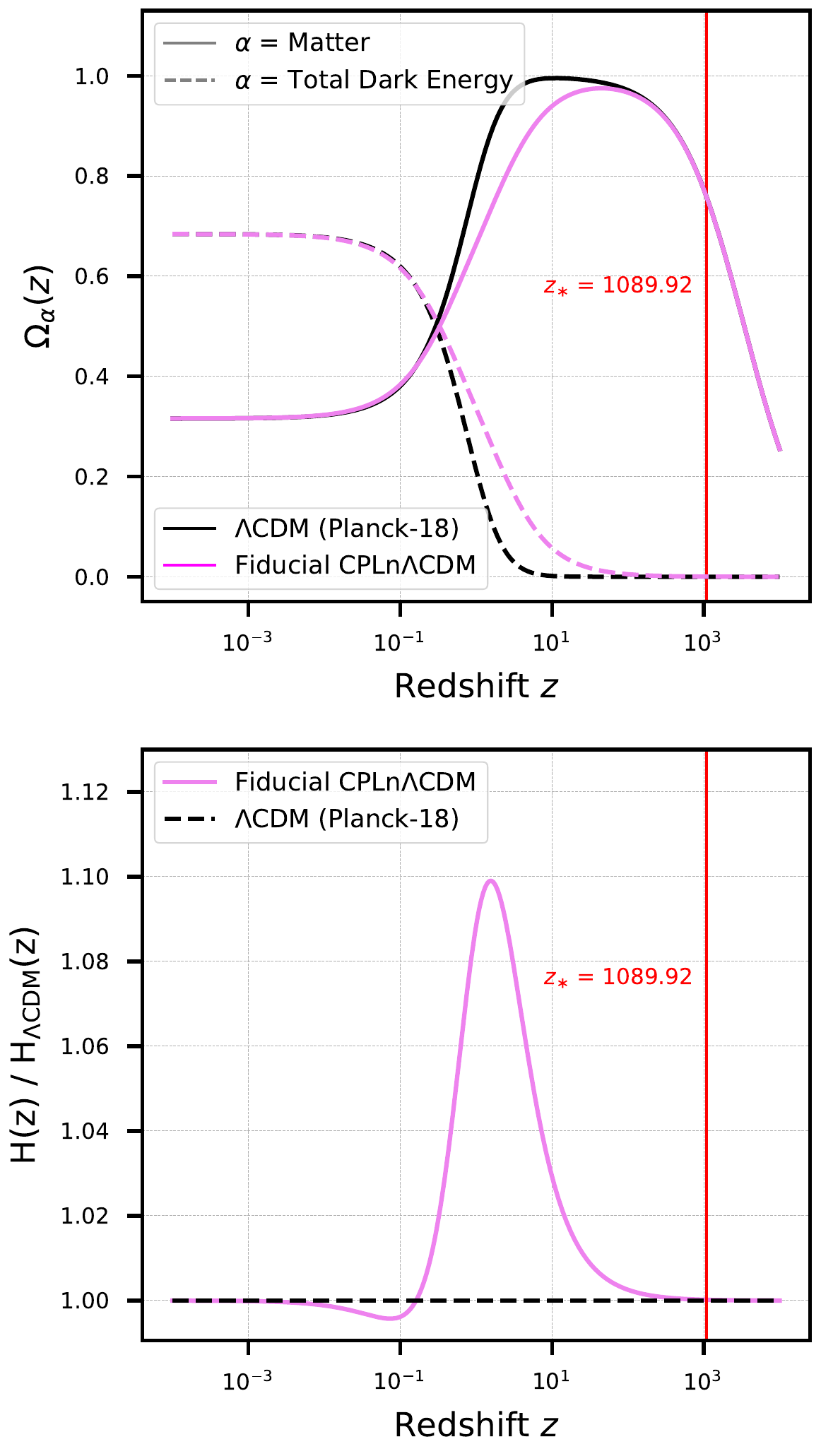}

\caption{Comparison of the evolution of energy densities and expansion rates across different cosmological models. The top panel shows the redshift evolution of the matter density (solid lines) and the total dark energy density (dashed lines). The bottom panel presents the ratio of the Hubble parameter in our fiducial CPL$n\Lambda$CDM model to that in the $\Lambda$CDM model, as a function of redshift.}
\label{fig:issues_engdens_exprate}
\end{figure}

While the analysis thus far may seem very promising and motivate a comprehensive MCMC study involving the simultaneous variation of both cosmological and astrophysical parameters to search for models in CPL$n\Lambda$CDM cosmology that can explain these high-$z$ galaxy observations, we pause to assess the broader viability of the \textbf{fiducial CPL$\bm{n\Lambda}$CDM} model beyond the high-redshift UVLFs and reionization datasets considered above — particularly in light of the wealth of high-precision cosmological observations that tightly constrain the evolutionary history of our Universe.

The temperature and polarization anisotropies observed in the CMB provide one of the most stringent tests of any cosmological model, as they encode detailed information about the Universe’s expansion history, composition, and geometry. The oscillatory features imprinted in the CMB power spectra define a characteristic angular scale $\theta_\ast$ on the sky, which is given by  $\theta_\ast = r_\ast/D_M(z_\ast)$ where $r_\ast$ is the comoving sound horizon at recombination and $D_M(z_\ast)$ is the comoving angular diameter distance to the recombination epoch at a redshift of $z_\ast$. This angular acoustic scale has been measured to an extremely high precision (0.03$\%$) by the latest observations of \citet{Planck2018} and is only weakly dependent on the cosmological model.

\begin{figure}
\centering
\includegraphics[width=\columnwidth]{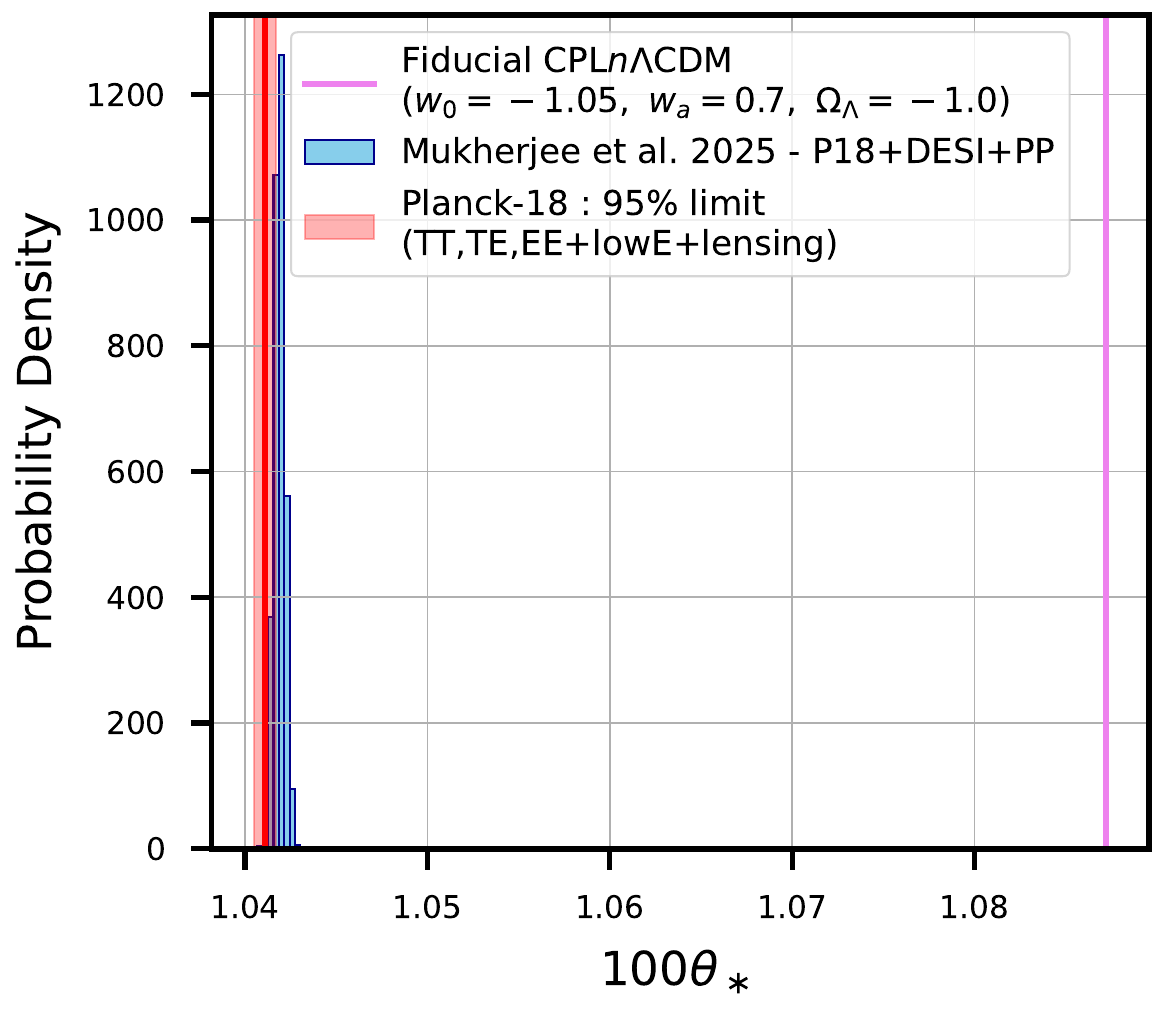}
\caption{The angular acoustic scale $\theta_\ast$ corresponding to acoustic oscillations imprinted in the CMB power spectra. The solid violet line indicates the value of $100\theta_*$ predicted by the \textbf{fiducial CPL$\bm{n\Lambda}$CDM} model (see \secn{subsec:results_fiducial_nCC_model}), while the red shaded region represents the corresponding 95\% confidence interval derived from the Planck-2018 analysis. The histogram in blue shows the distribution of $100 \theta_*$ for all CPL$n\Lambda$CDM models, that are consistent with other cosmological data such as CMB, BAO, SNe-Ia (\secn{subsec:results_maxboost_nCC_model}), taken from the MCMC chains of \citet[see \secn{subsec:results_maxboost_nCC_model}]{Mukherjee2025}.}
\label{fig:issues_thetaStar}
\end{figure}

Therefore, to assess the consistency of our fiducial CPL$n\Lambda$CDM scenario, it is instructive to examine how closely the angular acoustic scale $\theta_\ast$ predicted by the model agrees with these measurements. We present the evolution of the energy densities of the individual components, along with the Hubble expansion rate, for the \textbf{fiducial CPL$\bm{n\Lambda}$CDM} cosmological model in \fig{fig:issues_engdens_exprate}. Notably, both quantities converge to their $\Lambda$CDM counterparts prior to the epoch of recombination. Given that the size of the sound horizon at recombination $r_\ast$ is primarily determined by the speed of sound in the photon-baryon fluid, which in turn depends on the baryon-to-photon density ratio and the pre-recombination expansion history, we can therefore safely assume that its value remains unchanged in the \textbf{fiducial CPL$\bm{n\Lambda}$CDM} model. It is then straightforward to calculate $D_M(z_\ast)$ and thereafter, the value of $\theta_\ast$ for the cosmological model under consideration. However, as shown in \fig{fig:issues_thetaStar}, we find that the angular acoustic scale ($100\theta_*$) predicted by the \textbf{fiducial CPL$\bm{n\Lambda}$CDM} model is considerably larger than both the \textbf{$\bm{\Lambda}$CDM (Planck 2018)} constraints and the values obtained for CPL$n\Lambda$CDM models that fit a wide range of cosmological observations, including the CMB power spectra \citep[see \secn{subsec:results_maxboost_nCC_model}]{Mukherjee2025}. This indicates that while the \textbf{fiducial CPL$\bm{n\Lambda}$CDM} model shows promise by exhibiting better agreement with the high-redshift UVLF observations from JWST than our baseline \textbf{$\bm{\Lambda}$CDM (Planck 2018)} model, it is in strong tension with, and effectively ruled out by, other cosmological datasets -- in this case, the CMB itself. This highlights the importance of exercising caution when interpreting apparent successes of cosmological models in explaining isolated datasets. 

At first glance, one might be tempted to regard this discrepancy between the predicted and measured values of the acoustic scale $\theta_\ast$ as a consequence of holding $H_0$ fixed in our analysis, and to speculate that lowering $H_0$ -- which increases  $D_M(z_*)$ -- might restore consistency between the two datasets. However, such a line of thought is overly simplistic. CMB observations do not independently constrain $H_0$, $\Omega_m$, or $\Omega_b$, but instead tightly constrain the physical densities $\omega_m \equiv \Omega_m h^2$ and $\omega_b \equiv \Omega_b h^2$. Consequently, any change in $H_0$ that preserves consistency with the CMB must be accompanied by \textbf{correlated} changes in other cosmological parameters. For instance, lowering $H_0$ at fixed $\omega_m$ necessarily increases the present-day matter density parameter $\Omega_m$, thereby modifying the late-time expansion history, the growth of structures, and the late-time clustering amplitude ($\sigma_8$), which itself is anchored to the CMB constraints on the amplitude ($A_s$) and tilt ($n_s$) of the primordial power spectrum. Therefore, it is not entirely obvious that simply lowering $H_0$ would necessarily reconcile the CMB and JWST observations, unless the correlations among various cosmological parameters are treated in a fully self-consistent manner. Furthermore, even if apparent consistency with both datasets can be achieved, such parameter combinations could plausibly give rise to tensions with other late-time cosmological observables (see, e.g. \cite{Pedrotti2025}).

As a result, we now extend our analysis by incorporating constraints from other standard cosmological probes in order to identify more viable regions of parameter space within the CPL$n\Lambda$CDM cosmological framework through a systematic and self-consistent exploration in which all cosmological parameters are free to vary.


\subsection[]{A viable CPL$\bm{\lowercase{n}\Lambda}$CDM model exhibiting maximum boost in halo abundance (relative to $\bm{\Lambda}$CDM) at high redshifts ($\lowercase{\bm{z \approx 13. 2}}$)}
\label{subsec:results_maxboost_nCC_model}

Recently, \cite{Mukherjee2025}\footnote{\url{https://arxiv.org/abs/2501.18335v1}} carried out a comprehensive MCMC-based exploration of CPL$n\Lambda$CDM models that are compatible with a variety of cosmological observations — including CMB temperature, polarization, and lensing measurements from the Planck mission \citep{Planck2018}, baryon acoustic oscillations (BAO) data obtained by the Dark Energy Spectroscopic Instrument (DESI) Collaboration \citep{DESI2024_BAO}, and the Pantheon-Plus compilation of Type Ia supernovae light curves \citep{PantheonPlus_Brout2022}. We summarize the constraints obtained on the cosmological parameters from their analysis in \app{appendix:cosmoparams_P18+DESI+PP}.

\begin{table}
\centering
\caption{Cosmological parameters for the CPL$n\Lambda$CDM model that remains consistent with other cosmological observations while yielding the maximum enhancement in the abundance of dark matter haloes at $z = 13.2$, relative to the baseline \textbf{$\bm{\Lambda}$CDM (Planck 2018)} cosmological model. This model is referred to as the \textbf{maxboost CPL$\bm{n\Lambda}$CDM} model in the text. }
\begin{tabular}{|c|c|c|c|c|c|}
\hline
$\Omega_m$ & $\Omega_b h^2$ & $\Omega_\Lambda$ & $h$ & $\Omega_\phi$ \\
\hline
0.31012 & 0.0223509 & -0.117409 & 0.682049 & 0.807211 \\
\hline
 $w_0$ & $w_a$  & $n_s$ & $\sigma_8$ & $\tau_\mathrm{el}$ \\
\hline
 -0.887089 & -0.623657  & 0.966626 & 0.847624 & 0.0699301 \\
\hline
\end{tabular}
\label{tab:cosmo_params_maxboost}
\end{table}

\begin{figure*}
\centering
\includegraphics[width=\textwidth]{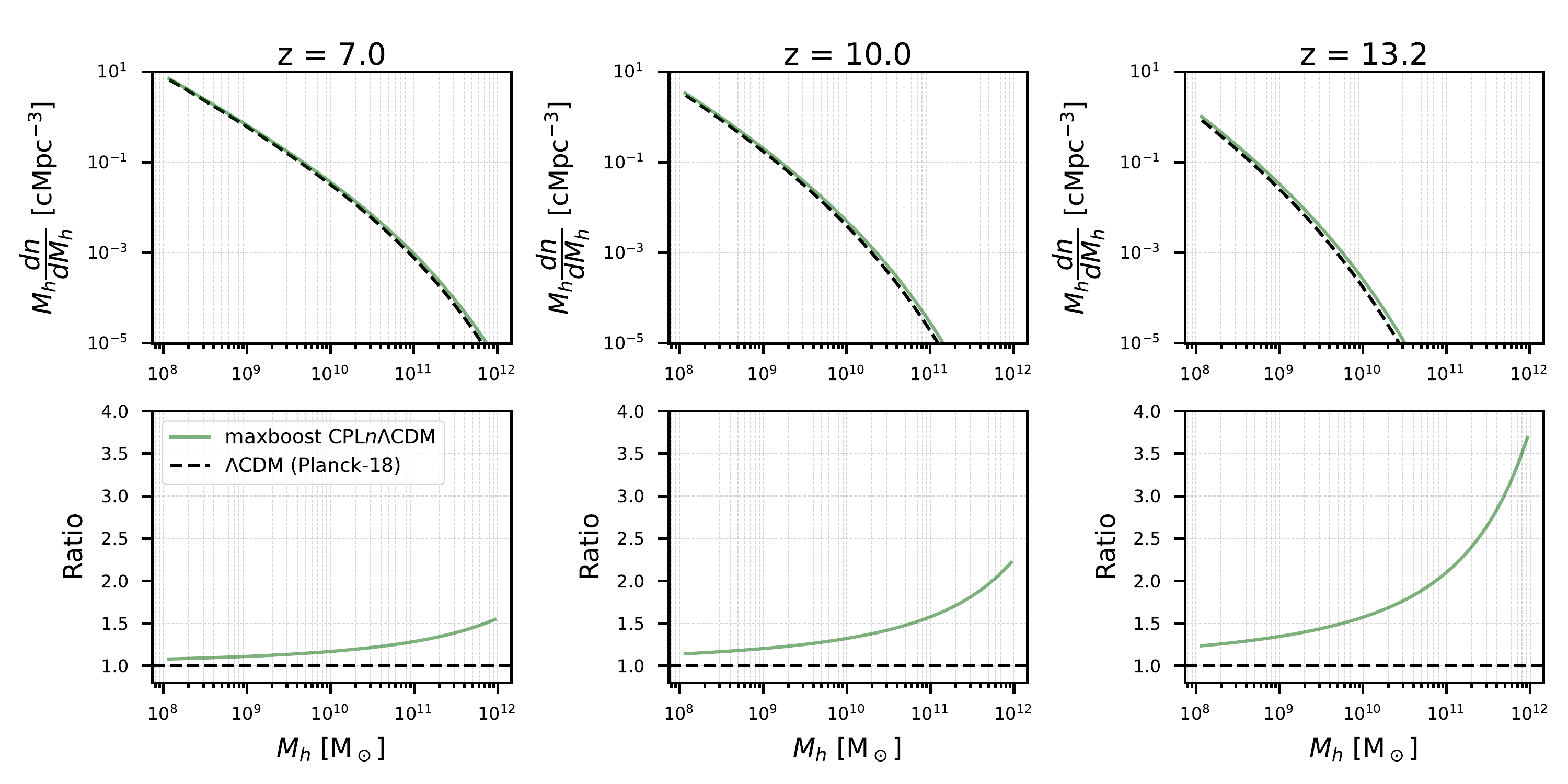}
\caption{Comparison of the dark matter halo mass functions at high redshifts ($z=7, 10,$ and $13.2$) for the \textbf{maxboost CPL$\bm{n\Lambda}$CDM} and the baseline \textbf{$\bm{\Lambda}$CDM (Planck 2018)} cosmological models. }
\label{fig:HMF_maxboost}
\end{figure*}

\begin{figure}
\centering
\includegraphics[width=\columnwidth]{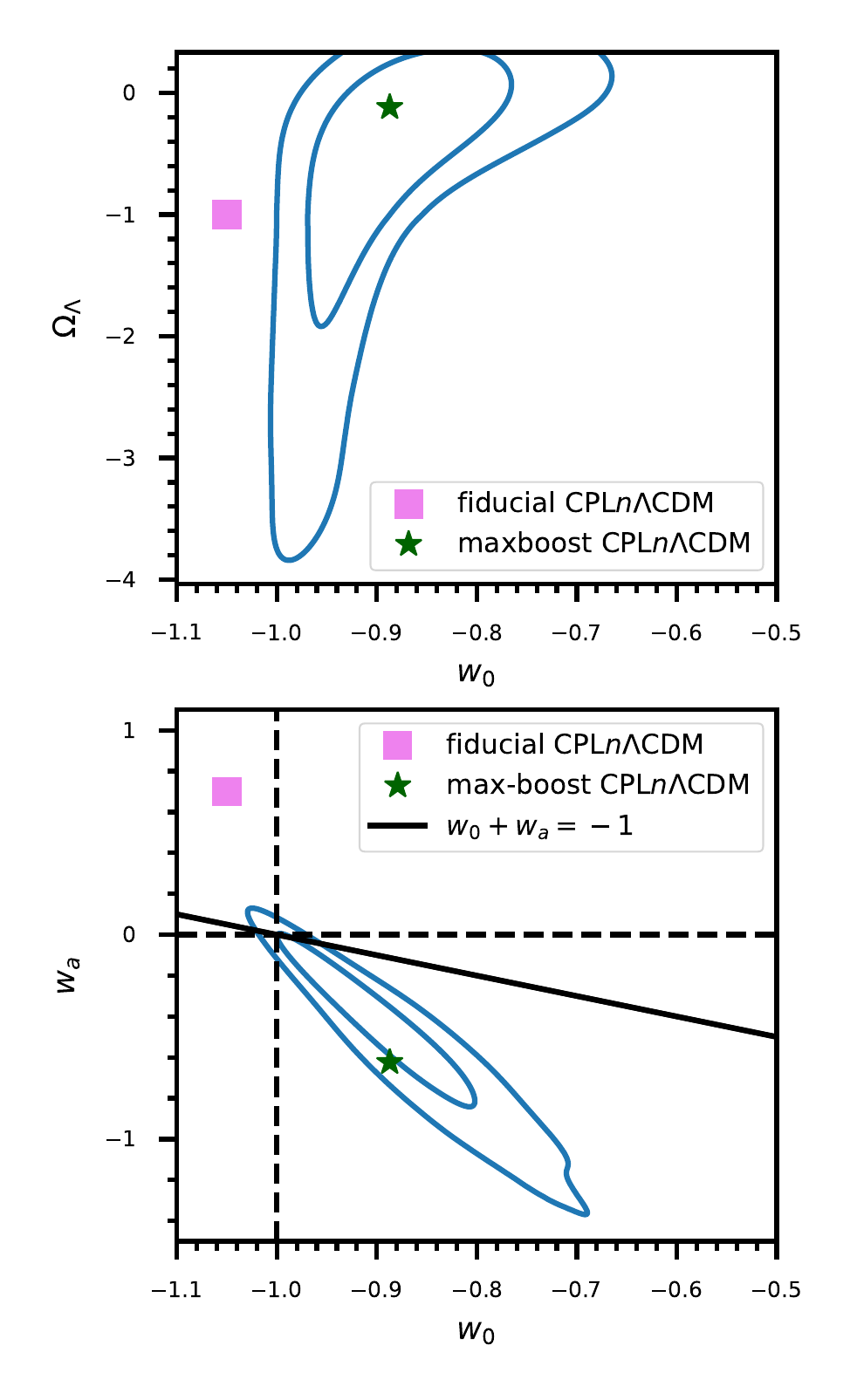}
\caption{The two-dimensional joint posterior distributions for some pairs of free parameters in the CPL$n\Lambda$CDM model obtained in \citet{Mukherjee2025} by comparing against the Planck-2018 CMB temperature, polarization, and lensing datasets, the DESI DR1 BAO measurements, and Pantheon-Plus compilation of Type-Ia supernovae light curves. The blue solid contours are drawn at 68\% and 95\% confidence levels. In each panel, the  \textbf{fiducial CPL$\bm{n\Lambda}$CDM}  and the \textbf{maxboost CPL$\bm{n\Lambda}$CDM} models are denoted using a violet square and a green star symbol, respectively.}
\label{fig:2D_contours}
\end{figure}

Instead of undertaking a complete exploration of the combined cosmological and astrophysical parameter space within the CPL$n\Lambda$CDM framework, we make use of samples from the posterior chains of \citet{Mukherjee2025} that are already consistent with several key cosmological datasets. This is sufficient to investigate the prospects of the ``allowed'' CPL$n\Lambda$CDM models in explaining the high-redshift galaxy and reionization observations.  For our analysis in this subsection, we discard the initial 30\% of the steps from their chains as burn-in and utilize the remaining samples.

From their chains, we identify the CPL$n\Lambda$CDM  model (i.e., with $\Omega_\Lambda<0$) that gives the maximum enhancement in the abundance of dark matter haloes at $z \approx 13.2$, relative to the baseline \textbf{$\bm{\Lambda}$CDM (Planck 2018)} cosmological model. We shall refer to this as the \textbf{``~maxboost CPL$\bm{n\Lambda}$CDM~''} model, whose corresponding cosmological parameters are listed in \tab{tab:cosmo_params_maxboost}. As seen in \fig{fig:HMF_maxboost}, this model yields a modest enhancement in halo number densities compared to the $\Lambda$CDM model — approximately a factor of 2 for 10$^{11} M_\odot$ haloes at $z = 13.2$. We further show this particular model within the marginalized two-dimensional joint posterior distribution space spanned by different parameter pairs in \fig{fig:2D_contours}. As noted in the previous section, it is not surprising that the \textbf{fiducial CPL$\bm{n\Lambda}$CDM} model (indicated by violet square symbols) lies outside the 95\% confidence region and is therefore ruled out at high significance.

\begin{figure*}
\centering
\includegraphics[width=\textwidth]{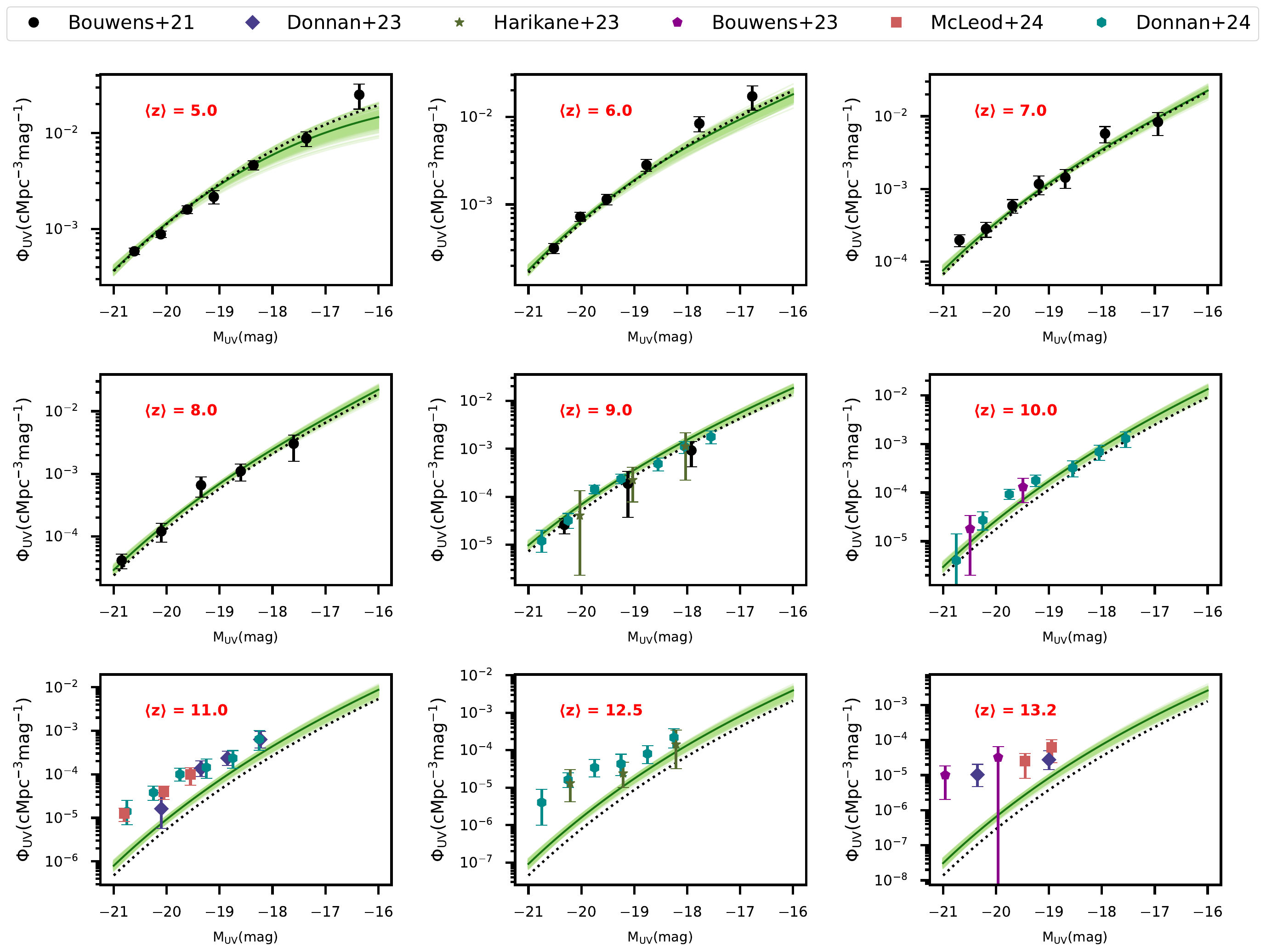}
\caption{The galaxy UV luminosity functions at nine different redshifts (with their respective mean values $\langle z \rangle$ mentioned in the upper left corner) for 200 random samples drawn from the MCMC chains of the \textbf{maxboost CPL$\bm{n\Lambda}$CDM} model. In each panel, the solid dark-green line corresponds to the best-fitting model, while the colored data points show the different observational constraints \citep{Bouwens2021, Donnan2023, Harikane2023, Bouwens2023, McLeod2024, Donnan2024} used in the likelihood analysis. The prediction from a model within the baseline \textbf{$\bm{\Lambda}$CDM (Planck 2018)} cosmology that best matches the observational measurements at $z < 10$ and does not assume any evolution in the UV efficiency parameters above $z \sim 10$ is also shown using black dotted lines.}
\label{fig:maxboost_UVLF}
\end{figure*}

\begin{figure*}
\centering
\includegraphics[width=\columnwidth]{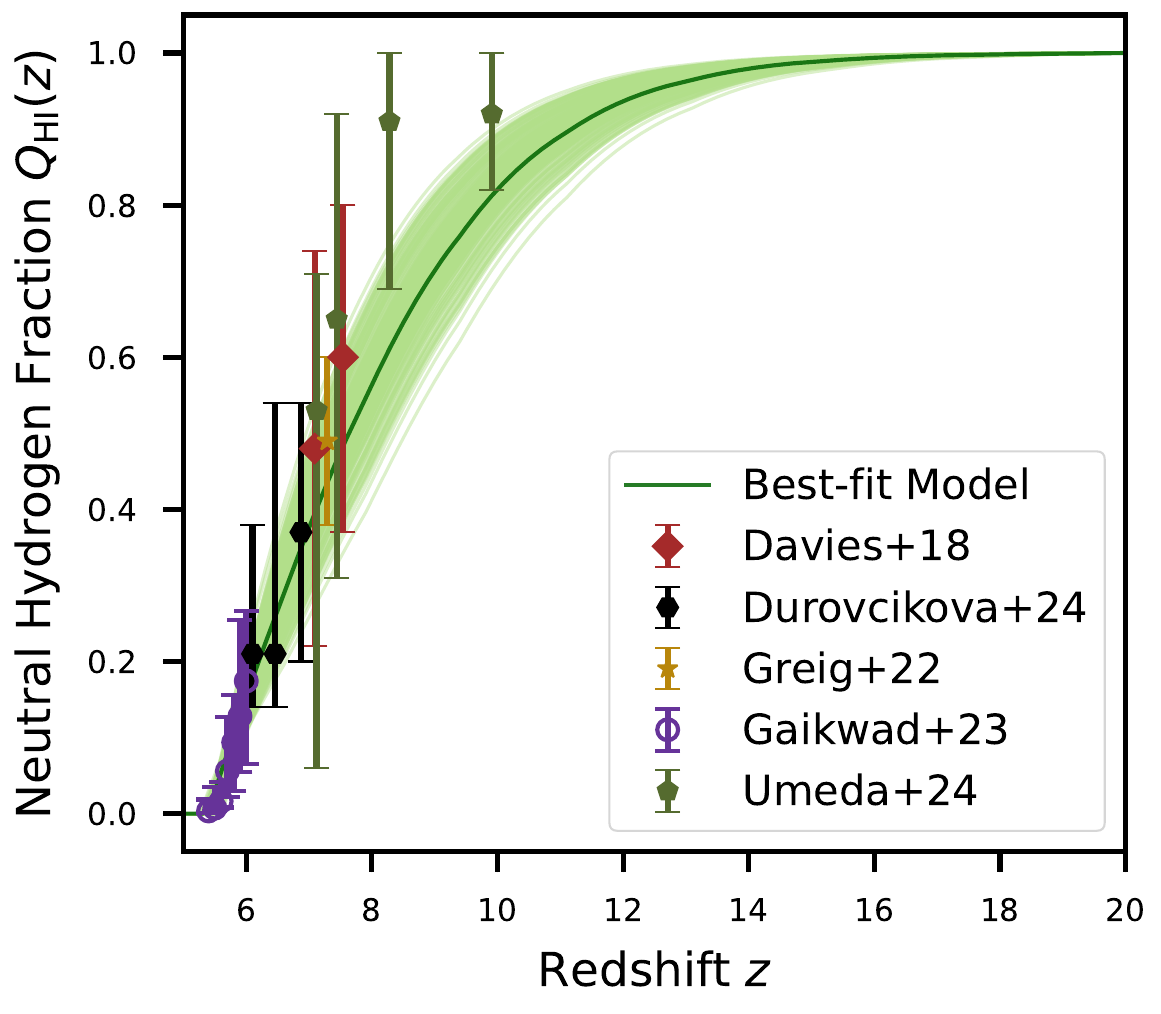}
\caption{The evolution of the globally averaged intergalactic neutral hydrogen fraction as a function of redshift for 200 random samples drawn from the MCMC chains of the  \textbf{maxboost CPL$\bm{n\Lambda}$CDM} model. The colored data points indicate the observational measurements of $Q_{\rm HI}$($z$) used in the analysis.} 
\label{fig:maxboost_reion_history}
\end{figure*}

As before, we perform an MCMC analysis to constrain the \textit{astrophysical} parameters of this model, under the assumption that they \textit{do not evolve with redshift}. For the likelihood calculations, we take the reionization optical depth corresponding to this model ($\tau_\mathrm{el} = 0.0699301$, listed in \tab{tab:cosmo_params_maxboost}) as the observational estimate. Its associated uncertainty is assumed to be the same as that obtained from the full MCMC chains (see \app{appendix:cosmoparams_P18+DESI+PP}), namely, $\Delta \tau_\mathrm{el} = 0.0073$.

The marginalized constraints on the free parameters are presented in the fourth column of \tab{tab:mcmc_results}. We show the model-predicted UVLFs for 200 random samples from the MCMC chains in \fig{fig:maxboost_UVLF}, and their corresponding reionization histories in \fig{fig:maxboost_reion_history}. We find that the UVLFs predicted by the \textbf{maxboost CPL$\bm{n\Lambda}$CDM} model are only marginally enhanced compared to those from the $\Lambda$CDM model. The predicted UVLFs are comparatively lower than the observational measurements of the bright-end ($\MUV \leq -20$) at $z \gtrsim 11$, with the discrepancy growing increasingly severe at even higher redshifts ($z \approx 13$) . By $z \approx 13$, the predicted UVLF is consistently lower than the observed data across the entire range of UV magnitudes probed by observations.  The model is, however, consistent with constraints on the progress of reionization as shown in \fig{fig:maxboost_reion_history}. As seen from \figs{fig:maxboost_reion_history}{fig:astroEvolution_reion_history}, reionization in the \textbf{maxboost CPL$\bm{n\Lambda}$CDM} model is more extended and starts earlier compared to the baseline \textbf{$\bm{\Lambda}$CDM (Planck 2018)} cosmological model, thereby yielding a larger value of $\tau_{\rm el}$. However, the mean electron scattering optical depth derived from our MCMC analysis ($\tau_\mathrm{el} \approx 0.0597$) is still $\approx 1.3 \sigma$ \footnote{Here, the $\sigma$ has been estimated using the quadrature sum of the two individual `symmetrized' error uncertainties} lower than the value ($\tau_\mathrm{el} \approx 0.0699$) actually corresponding to this cosmological model. This may be attributed to the inclusion of $Q_{\mathrm{HI}}$ measurements (without any correction, from the different studies listed in \secn{sec:data_and_methods}) in the likelihood calculation, which prevents reionization histories with arbitrarily large values of $\tau_{\rm el}$ from being accepted while scanning the parameter space. The escaping ionizing efficiency exhibits a weak dependence on halo mass, characterized by a marginally negative power-law slope of $\alpha_{\rm esc} = -0.087^{+0.063}_{-0.072}$. Interestingly, the critical halo mass $\Mcrit$ affected by radiative feedback is found to be approximately $10^{10.5} M_\odot$, which is slightly higher than that obtained within the baseline \textbf{$\bm{\Lambda}$CDM (Planck 2018)} model (\app{appendix:LCDM_astro_evolution}). This higher threshold not only facilitates the completion of reionization by $z \approx 5$, but also enables the model to remain consistent with the observed UVLFs at lower redshifts ($z < 7$, when radiative feedback is dominant with reionization nearing completion) so that the redshift-independent UV efficiency parameter, $\varepsilon_\mathrm{*10,UV}$, can be flexibly calibrated to reproduce the UVLF observations across the entire redshift range of $5 \leq z < 14$. 

Therefore, our analysis reveals that the \textbf{maxboost CPL$\bm{n\Lambda}$CDM} cosmological model struggles to simultaneously reproduce the full shape and evolution of the UVLFs across the redshift range $5 \leq z < 14$, when assuming redshift-independent astrophysical properties for high-$z$ galaxies. Our findings lend support to the emerging consensus in recent literature that cosmological modifications \textbf{\textit{alone}} are likely insufficient to account for the unexpectedly high abundance of $z > 10$ galaxies observed by JWST. For instance, \citet{Sabti2024} recently concluded that alterations to the $\Lambda$CDM matter power spectrum (and thereby, leading to an enhancement of the halo mass function) that is large enough to reproduce the abundance of $z>10$ JWST candidates would lead to inconsistencies with the UVLFs measured by HST at lower redshifts and/or other cosmological datasets. Similarly, \citet{Shen2024} showed that while cosmological models incorporating early dark energy (EDE) can successfully reproduce the UVLFs in the range $4 \lesssim z \lesssim 10$, they still require either a modest increase in star formation efficiency or a nominal degree of stochasticity in UV emission from galaxies to match the JWST observations at higher redshifts - $12 \lesssim z \lesssim 16$. A similar conclusion was also reached by \citet{Liu2024}, who investigated the prospects of explaining the JWST observations using EDE models and found that these models can match the UVLFs at $z > 10$ better than $\Lambda$CDM but don't fare as well as $\Lambda$CDM at lower redshifts ($z < 10$). 

Given that even the CPL$n\Lambda$CDM model, which yields the maximal halo abundance at $z \approx 13$, fails to consistently reproduce the observed evolution of the UV luminosity function from $z = 5$ to $z = 13.2$, we also examined whether a model that maximizes the ratio of the boost in halo abundances (relative to the baseline \textbf{$\bm{\Lambda}$CDM (Planck 2018)} model) between $z = 13.2$ and $z = 5$ can provide a better match (see \app{subsec:results_relative-zboost_nCC_model} for details). We find that our conclusions remain unchanged, underscoring the difficulty that CPL$n\Lambda$CDM cosmological models face in simultaneously reproducing the galaxy UV luminosity function across the entire redshift range from $z = 13.2$ to $z = 5$, without any astrophysical evolution.

We should mention here that some intriguing proposals have been explored in the literature that argue in favor of cosmological solutions for the JWST galaxy excess. For instance, \citet{Padmanabhan2023} investigated how a modified matter transfer function could enhance early structure formation, while \citet{Menci2024} examined models incorporating a negative cosmological constant, similar in spirit to the framework we use here. Our analysis is intended to build upon this line of inquiry by focusing on a specific and crucial question: to what extent can such models enhance the early halo abundance while remaining simultaneously consistent with the full suite of precision cosmological datasets (e.g., CMB, BAO, SNe)? This focus is motivated by our finding in this section that some promising high-redshift models can face challenges when confronted with these other stringent constraints. By carefully selecting for CPL$n\Lambda$CDM models that satisfy this holistic viability, our work quantifies the level of enhancement permitted by the existing cosmological data. Our finding that even these models fall short of fully explaining the observations strengthens the case that modifications to astrophysics are a key part of the complete solution.

\section{Conclusion}
\label{sec:conclusion}

The James Webb Space Telescope (JWST) is revolutionizing our view of the early universe by revealing an unexpected abundance of luminous galaxies at redshifts $z > 10$ \citep{Naidu2022, Castellano2022, Finkelstein2022, Labbe2023, Atek2023, Adams2023, Bradley2023, Whitler2023, Robertson2024, Castellano2023_UVLF, Finkelstein2023_UVLF, Gonzalez2023, Donnan2023, Harikane2023, Bouwens2023, McLeod2024, Adams2024, Finkelstein2024_UVLF, Whitler2025, Gonzalez2025, Weibel2025}. The existence of these ultra-luminous galaxies at very early cosmic epochs challenges the standard cosmological model, $\Lambda$CDM, prompting a critical re-evaluation of the intertwined physics of cosmic expansion and early galaxy formation. This work confronts a key question: Can modifications to cosmology \textbf{\textit{alone}} account for these surprising observations, or are changes to our understanding of galaxy formation physics inevitable?

We investigate a promising class of alternative models (CPL$n\Lambda$CDM) featuring dynamical dark energy and a negative cosmological constant. Using a self-consistent framework that couples galaxy evolution with cosmic reionization \citep{Chakraborty2024, Chakraborty2026}, we test whether such a model, itself constrained by established cosmological data like the CMB, can reproduce the observed galaxy populations from $z \approx 5$ to $z \approx 14$ \textit{without invoking any evolution in their astrophysical properties}. 

Our main conclusions are:
\begin{itemize}
    \item \textbf{A purely cosmological enhancement is insufficient.} 
    While CPL$n\Lambda$CDM models that are consistent with other cosmological probes perform slightly better than the standard $\Lambda$CDM scenario at $z>10$, they fail to reproduce the full shape and evolution of the observed galaxy UV luminosity function across the redshift range $5 \leq z < 14$. They cannot, on their own, resolve the tension. This suggests that some degree of evolution in the astrophysical properties of high-redshift galaxies -- albeit less pronounced -- may still be necessary.
    
    \item \textbf{High-redshift observations must be reconciled with the established cosmic history.} 
    Our analysis serves as a crucial case study, demonstrating that any proposed cosmological solution to the early galaxy excess must be rigorously tested against the full suite of precision cosmological observations. Apparent successes in one observational regime can be decisively ruled out by constraints from another.
\end{itemize}

Our findings contribute to a growing consensus that the surprising abundance of early, bright galaxies is unlikely to be resolved by modifications to the cosmological model alone \citep{Sabti2024, Gouttenoire2024, Davari2024} and that even under modified cosmological scenarios, an evolution in galaxy properties is still required \citep{Shen2024, Liu2024, Ellis2025}. This places renewed emphasis on understanding the astrophysical processes, such as star formation efficiency, feedback, or a top-heavy stellar initial mass function, that govern galaxy formation in the first billion years. Assuming plausible ionizing properties for high-redshift galaxies, CPL$n\Lambda$CDM models also produce a reionization history consistent with constraints on the ionization state of the intergalactic medium derived from astrophysical observations. We note, however, that these observational constraints on the reionization timeline available in the literature have not been explicitly evaluated for alternative cosmologies such as the one considered in this work.

Perhaps more importantly, this work demonstrates a powerful synergy. We show that the available datasets of high-redshift galaxies are not only probes of astrophysics but can also serve as a potent new tool to constrain fundamental cosmology. By systematically marginalizing over uncertain galaxy properties, we turn the challenge posed by JWST into a novel opportunity to test the viability of beyond-$\Lambda$CDM models.

A crucial next step is to move beyond the halo mass function and model the spatial distribution of galaxies and the topology of reionization within these alternative cosmologies. While full, high-resolution hydrodynamical simulations are an important long-term goal, their computational expense makes them impractical for exploring the vast and degenerate parameter space where both cosmological and astrophysical properties vary. A more immediate and powerful path forward lies in combining our analytical framework with efficient, large-volume semi-numerical codes.

To this end, we plan to integrate our model with \texttt{SCRIPT}, a semi-numerical model of galaxy formation and reionization developed by some of the authors \citep{Choudhury2018, Choudhury2025}. This hybrid approach will enable us to self-consistently generate a rich set of mock observables, including galaxy luminosity functions, clustering statistics, the Lyman-$\alpha$ forest opacity, and the 21\,cm signal, for a wide range of CPL$n\Lambda$CDM and astrophysical scenarios. Confronting these detailed predictions with the wealth of data from JWST, and soon from the Roman Space Telescope and 21\,cm experiments like the SKA, will be indispensable for decisively disentangling the signatures of new physics from the complexities of galaxy formation at cosmic dawn.


\section*{Acknowledgments}

AC and TRC acknowledge support from the Department of Atomic Energy, Government of India, under project no. 12-R\&D-TFR-5.02-0700.  AAS acknowledges funding from Anusandhan National Research Foundation (ANRF), Government of India, under the research grant no. CRG/2023/003984. PM acknowledges funding from the ANRF, Government of India, under the National Post-Doctoral Fellowship (File no. PDF/2023/001986).

\section*{Data Availability}

The data generated and presented in this paper will be made available upon reasonable request to the corresponding author (AC).



\bibliographystyle{mnras}
\bibliography{mnras_ver2} 



\newpage


\appendix
\section{Constraints on the redshift evolution of galaxy properties from JWST UVLF observations in the $\Lambda$CDM cosmological framework}
\label{appendix:LCDM_astro_evolution}

In this appendix, we summarize the constraints obtained on the evolving astrophysical properties of high-$z$ galaxies from a Bayesian analysis using the observational datasets discussed in \secn{sec:data_and_methods}. Here, the background cosmological model is assumed to be \textbf{$\bm{\Lambda}$CDM (Planck 2018)}, with the various cosmological parameters set to the values mentioned at the end of \secn{sec:intro}. This model shall be referred to as the \textbf{$\bm{\Lambda}$CDM+astro-evolution} model.

The astrophysical parameters $\log_{10}(\varepsilon_{\rm *10,UV})$ and $\alpha_*$ in this case are allowed to evolve with redshift following the flexible tanh parameterization introduced in our earlier works \citepalias{Chakraborty2024, Chakraborty2026}, which is given by --   
\begin{eqnarray}
\log_{10}\varepsilon_{\rm *10,UV}(z) = \dfrac{\ell_{\varepsilon, \mathrm{hi}} + \ell_{\varepsilon, \mathrm{lo}}}{2} + \dfrac{\ell_{\varepsilon, \mathrm{hi}} - \ell_{\varepsilon, \mathrm{lo}}}{2} \tanh\left(\dfrac{z-z_\ast}{\Delta z_\ast}\right),
\end{eqnarray}
and
\begin{eqnarray}
\alpha_\ast(z) = \dfrac{\alpha_{\mathrm{hi}} + \alpha_{\mathrm{lo}} }{2}  + \dfrac{\alpha_{\mathrm{hi}} - \alpha_{\mathrm{lo}}}{2} \tanh\left(\dfrac{z-z_\ast}{\Delta z_\ast}\right).
\end{eqnarray}

Under this parameterization, the parameter $\log_{10} \varepsilon_{*10,\mathrm{UV}} $ ($\alpha_\ast$) evolves from $\ell_{\varepsilon, \mathrm{lo}}$  ($\alpha_{\mathrm{lo}}$) at low redshifts to $\ell_{\varepsilon, \mathrm{hi}}$  ($\alpha_{\mathrm{hi}}$) at high redshifts, with the transition centred at a characteristic redshift $z_\ast$ and extending over a redshift range $\Delta z_\ast$. 

\begin{figure}
\centering
\includegraphics[width=\columnwidth]{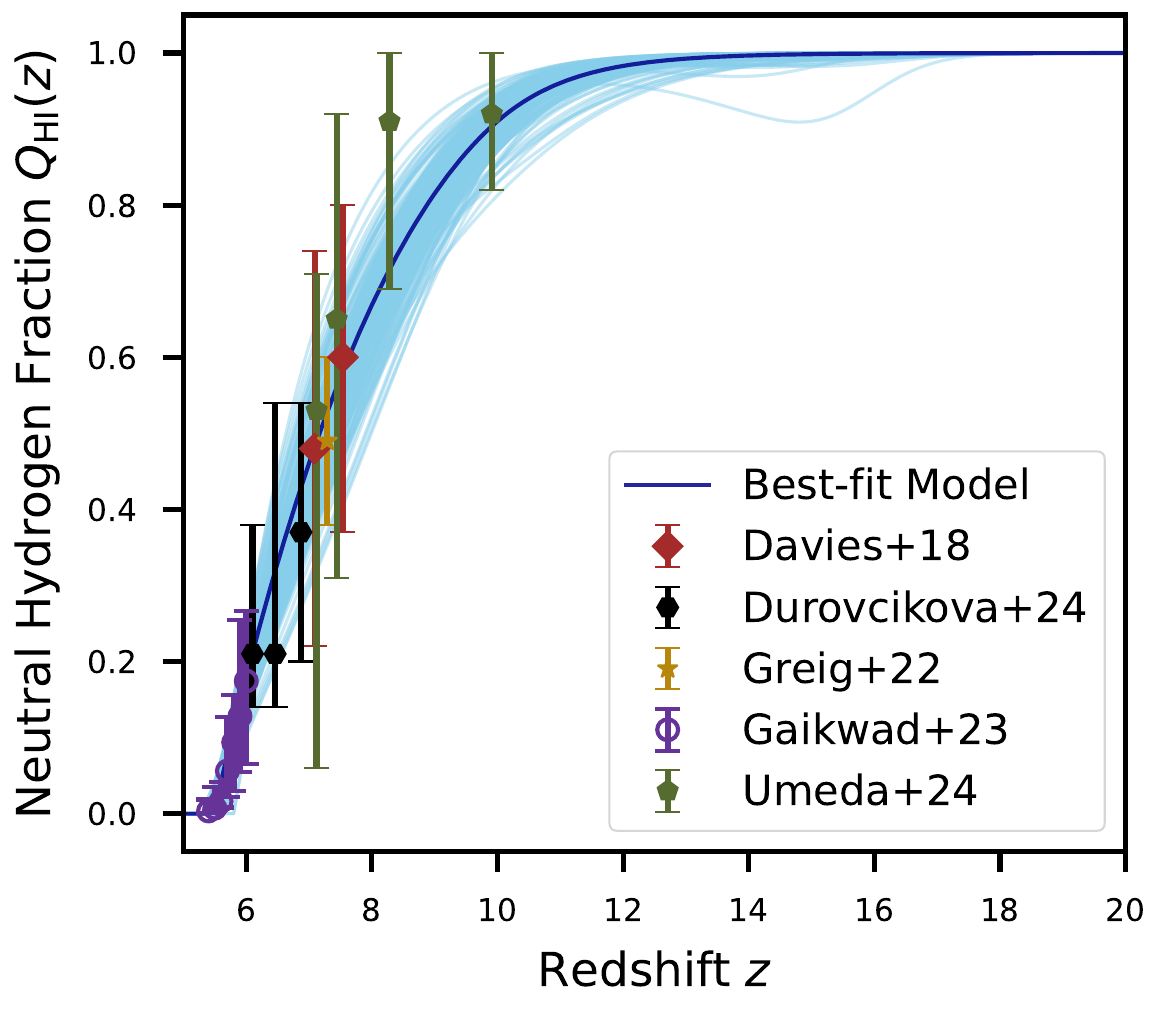}
\caption{The evolution of the globally averaged intergalactic neutral hydrogen fraction as a function of redshift for 200 random samples drawn from the MCMC chains of the \textbf{$\bm{\Lambda}$CDM+astro-evolution} model. The colored data points indicate the observational measurements of $Q_{\rm HI}$($z$) used in the analysis.} 
\label{fig:astroEvolution_reion_history}
\end{figure}

The constraints on the nine free parameters characterizing the astrophysical properties of galaxies are mentioned in \tab{tab:LCDM_astro_evolve_results}. The implications of these constraints have been discussed extensively in our earlier papers. We show the reionization history for 200 random samples drawn from the MCMC chains of \textbf{$\bm{\Lambda}$CDM+astro-evolution} model in \fig{fig:astroEvolution_reion_history} to serve as a benchmark for understanding how changes in the background cosmological model affect the global reionization process.

\begin{table}
 \caption{Parameter constraints obtained from the MCMC-based analysis. The first nine rows correspond to the free parameters of the \textbf{$\bm{\Lambda}$CDM+astro-evolution} model, while the remaining are the derived parameters. The free parameters are assumed to have uniform priors in the range mentioned in the second column. The numbers in the last column shows the mean value with 1$\sigma$ errors for different parameters of the model.\\} 
 \label{tab:LCDM_astro_evolve_results}
 \begin{tabular*}{\columnwidth}{c@{\hspace*{33pt}}c@{\hspace*{33pt}}c@{\hspace*{33pt}}}
 \hline   \hline  \\
    Parameters & Priors &  68\% limits \\  \\
  \hline   \hline \\
{\boldmath$\ell_{\varepsilon,\mathrm{hi}}$} & [-2.0, 2.0] &  $-0.068^{-0.044}_{-0.48}   $\\ \\

{\boldmath$\ell_{\varepsilon,\mathrm{lo}}$} & [-2.0, 0.5] & $-0.889^{+0.070}_{-0.045}  $ \\ \\

{\boldmath$z_\ast $} & [8.0, 18.0] & $11.84^{+0.38}_{-1.5} $ \\ \\

{\boldmath$\Delta z_\ast $} & [0.5, 6.0] & $1.89^{+0.39}_{-0.94}$\\ \\

{\boldmath$\alpha_\mathrm{hi}$} & [0.0, 7.0] & $2.11^{+0.15}_{-1.5} $  \\ \\

{\boldmath$\alpha_\mathrm{lo}$} & [-1.0, 1.0]  & $0.302^{+0.036}_{-0.050}   $  \\ \\

{\boldmath$\log_{10}~(\varepsilon_{\mathrm{esc,10}})$} & [-3.0, 1.0]  & $-0.810\pm 0.043$  \\ \\

{\boldmath$\alpha_{\rm esc}$} & [-3.0, 1.0] & $-0.17^{+0.14}_{-0.11}    $  \\ \\

{\boldmath$\log_{10}(M_{\mathrm{crit}}/M_\odot)$} & [9.0, 11.0] &$10.11^{+0.28}_{-0.11}    $ \\ \\

\hline \\

$\tau_{\rm el}  $ & - &  $0.0540^{+0.0020}_{-0.0023}$ \\ \\

$\ell_{\varepsilon,\mathrm{hi}} - \ell_{\varepsilon,\mathrm{lo}} $ & -  &$0.822^{-0.014}_{-0.52}$  \\ \\

$\alpha_\mathrm{hi}  - \alpha_\mathrm{lo}  $ & - & $1.81^{+0.15}_{-1.5}$\\ \\
\hline
\end{tabular*}
\end{table}

\newpage


\section[]{Constraints on the cosmological parameters in the CPL$\bm{\lowercase{n}\Lambda}$CDM cosmology, as obtained by Mukherjee et al. (2025)}
\label{appendix:cosmoparams_P18+DESI+PP}

In this appendix, we list the constraints (at 68\% confidence level) obtained by \cite{Mukherjee2025}\footnote{\url{https://arxiv.org/abs/2501.18335v1}} on the cosmological parameters of the CPL$n\Lambda$CDM model from a joint analysis of the full Planck-PR3 CMB TT, EE, TE and lensing data \citep{Planck2018}, DESI-BAO Data Release 1 measurements \citep{DESI2024_BAO}, and the Pantheon Plus compilation of SN-Ia light curves \citep{PantheonPlus_Brout2022}. Here, we have removed the first $30\%$ of the samples from their chains as burn-in and used the remaining samples to derive these constraints. 
\begin{table}
\centering
\caption{Constraints on the cosmological model parameters obtained by jointly analyzing the full Planck-2018 CMB dataset (temperature, polarization, and lensing), the DESI DR1 BAO measurements, and the latest Pantheon-Plus compilation of light curves of spectroscopically confirmed Type Ia supernovae.}
\begin{tabular}{c c}
\hline\hline
\multicolumn{2}{c}{\rule{0pt}{4ex}Planck-18 + DESI-DR1 + Pantheon-Plus } \\[.5ex]
\hline\hline \\
\textbf{Sampled Parameters} & 68\% limits \\[0.5ex]
\hline \\
{\boldmath$\log(10^{10} A_\mathrm{s})$} & $3.041\pm 0.014$ \\ \\
{\boldmath$n_\mathrm{s}$} & $0.9657\pm 0.0038$ \\ \\
{\boldmath$100\theta_\mathrm{s}$} & $1.04190\pm 0.00029$ \\ \\
{\boldmath$\Omega_\mathrm{b} h^2$} & $0.02240\pm 0.00014$ \\ \\
{\boldmath$\Omega_\mathrm{c} h^2$} & $0.11948\pm 0.00099$ \\ \\
{\boldmath$\tau_\mathrm{reion}$} & $0.0536\pm 0.0073$ \\ \\
{\boldmath$w_0$} & $-0.887^{+0.039}_{-0.087}$ \\ \\
{\boldmath$w_a$} & $-0.47^{+0.37}_{-0.15}$ \\ \\
{\boldmath$\Omega_\phi$} & $< 1.41$ \\[0.5ex]
\hline \\
\textbf{Derived Parameters} & 68\% limits \\[0.5ex]
\hline \\
$A_\mathrm{s}$ & $\left(2.093\pm 0.030\right)\times 10^{-9}$ \\ \\
$H_0$ & $68.03\pm 0.73$ \\ \\
$\Omega_\mathrm{m}$ & $0.3081\pm 0.0068$ \\ \\
$\Omega_\Lambda$ & $-0.680^{+0.78}_{-0.058}$ \\ \\
$\sigma_8$ & $0.816\pm 0.010$ \\ \\
$S_8$ & $0.827\pm 0.010$ \\ \\
$r_\mathrm{drag}$ & $147.21\pm 0.24$ \\ \\
$\Omega_b$ & $0.0484\pm 0.0011$ \\ \\[0.5ex]
\hline\hline
\end{tabular}
\label{tab:Mukherjee2025_constraints}
\end{table}

\newpage


\section[]{A  CPL$\bm{\lowercase{n}\Lambda}$CDM model that maximizes the boost in halo abundance \lowercase{(relative to} $\bm{\Lambda}$CDM) at $\lowercase{\bm{z \approx 13. 2}}$ compared to $\lowercase{\bm{z \approx 5}}$}
\label{subsec:results_relative-zboost_nCC_model}

\begin{figure*}
\centering
\includegraphics[width=\textwidth]{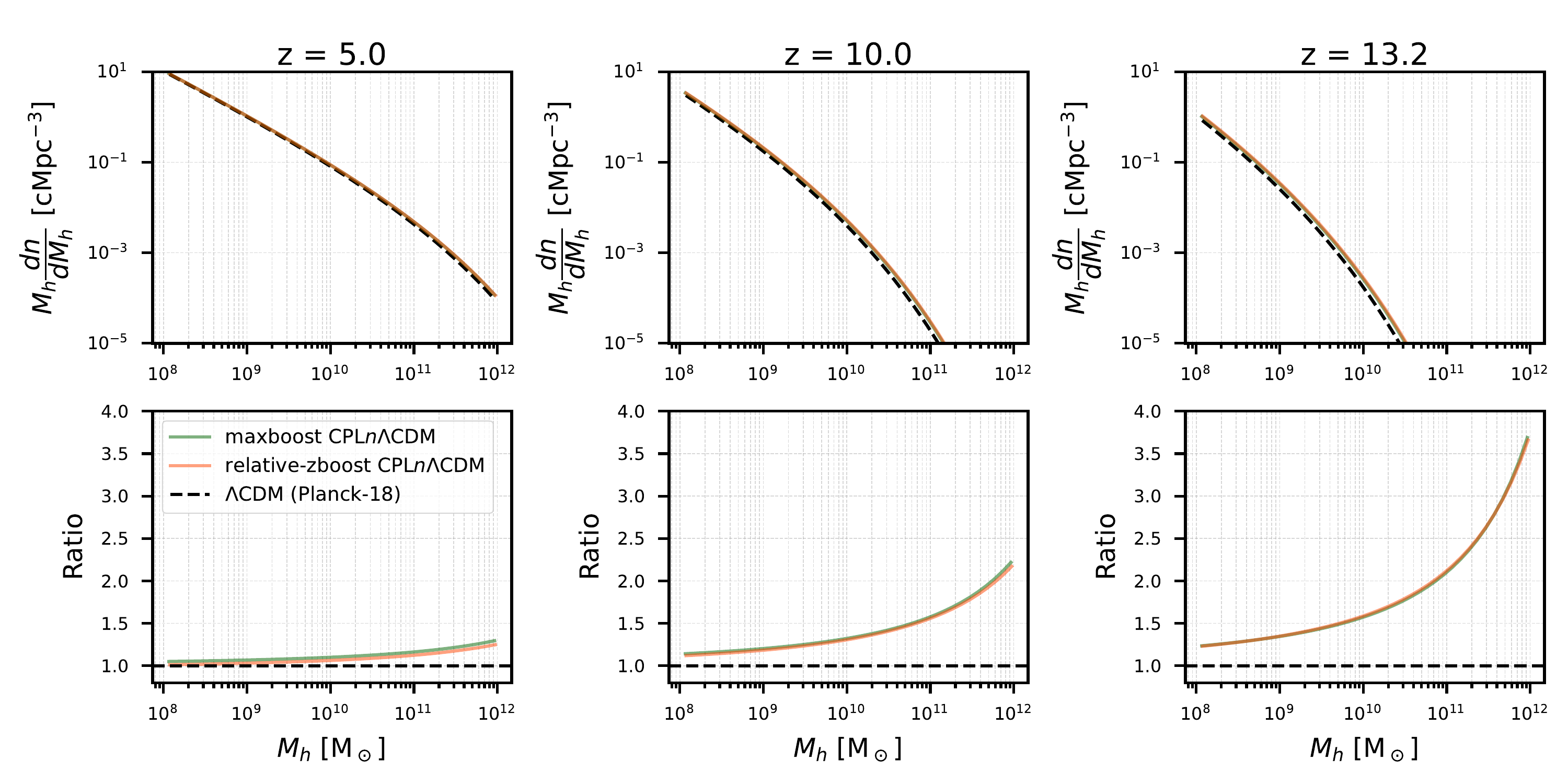}
\caption{Comparison of the dark matter halo mass functions at high redshifts ($z=5, 10,$ and $13.2$) within the \textbf{maxboost CPL$\bm{n\Lambda}$CDM} and \textbf{relative-zboost CPL$\bm{n\Lambda}$CDM}  models with that obtained for the baseline \textbf{$\bm{\Lambda}$CDM (Planck 2018)} cosmological model. }
\label{fig:HMF_max_and_relative_boost}
\end{figure*}

\begin{figure*}
\centering
\includegraphics[width=\textwidth]{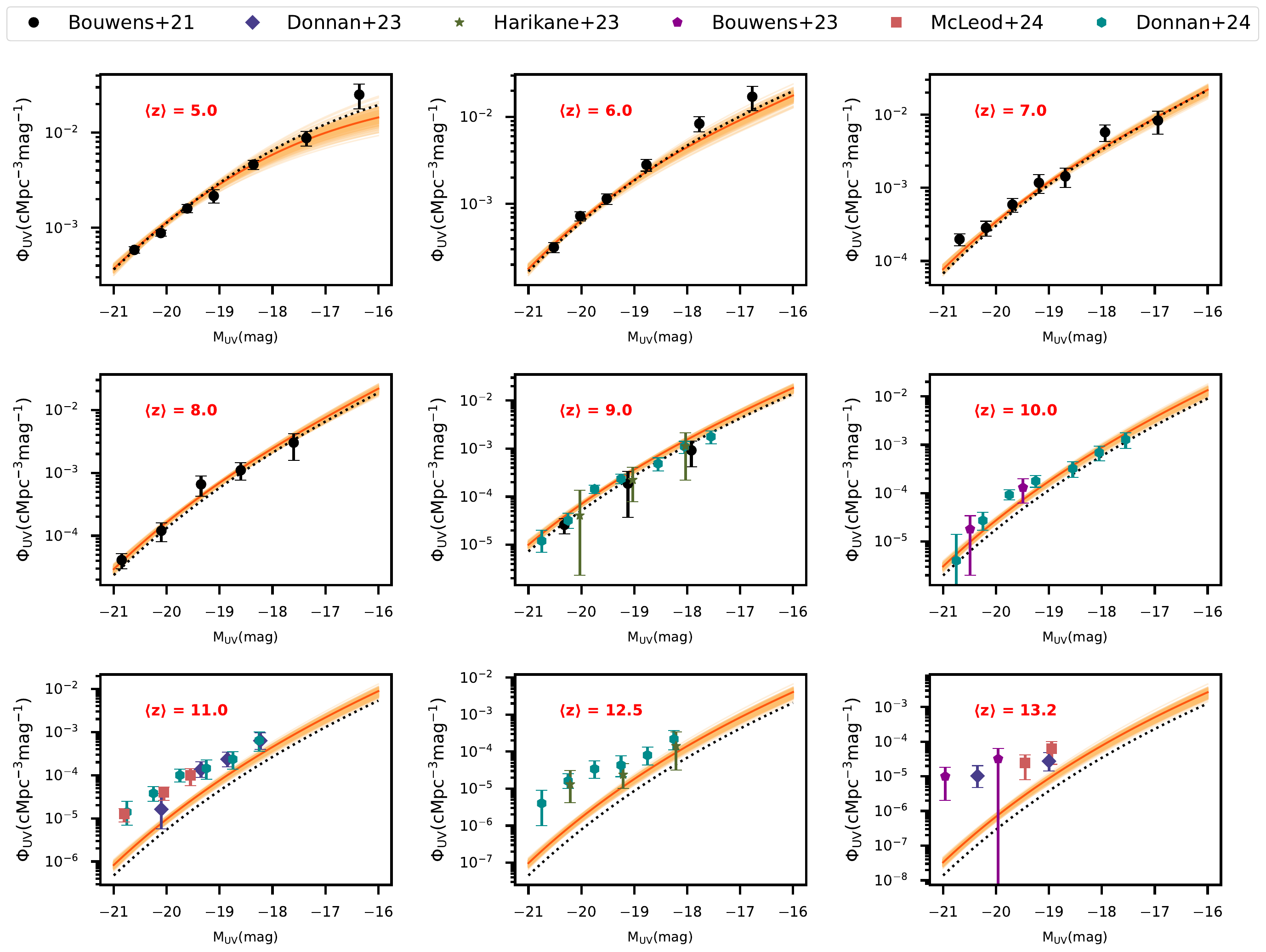}
\caption{The galaxy UV luminosity functions at nine different redshifts (with their respective mean values $\langle z \rangle$ mentioned in the upper left corner) for 200 random samples drawn from the MCMC chains of the \textbf{relative-zboost CPL$\bm{n\Lambda}$CDM} model. In each panel, the solid dark-orange line corresponds to the best-fitting model, while the colored data points show the different observational constraints \citep{Bouwens2021, Donnan2023, Harikane2023, Bouwens2023, McLeod2024, Donnan2024} used in the likelihood analysis. The prediction from a model within the \textbf{$\bm{\Lambda}$CDM (Planck-2018)} cosmology that best matches the observational measurements at $z < 10$ and does not assume any evolution in the UV efficiency parameters above $z \sim 10$ is also shown using black dotted lines.}
\label{fig:relative-zboost_UVLF}
\end{figure*}

In this appendix, we carry out an MCMC analysis similar to that done in \secn{subsec:results_maxboost_nCC_model}, except that we now select the CPL$n\Lambda$CDM  cosmological model that maximizes the ratio of the enhancement in the abundance of dark matter haloes of fixed mass (relative to the baseline \textbf{$\bm{\Lambda}$CDM (Planck 2018)} model) at $z \approx 13.2$ to that at $z = 5$. We shall refer to this as the \textbf{``~relative-zboost CPL$\bm{n\Lambda}$CDM~''} model, whose corresponding cosmological parameters are listed in \tab{tab:cosmo_params_max_relative-zboost}. The enhancement in halo number densities relative to the $\Lambda$CDM model for this scenario is presented in \fig{fig:HMF_max_and_relative_boost}. The corresponding UV luminosity functions, derived for 200 random samples from the MCMC analysis, are shown in \fig{fig:relative-zboost_UVLF}. As evident from these figures, we find the results for this analysis to be qualitatively similar to those obtained in \secn{subsec:results_maxboost_nCC_model}, thereby raising considerable apprehension on the viability of CPL$n\Lambda$CDM models as an explanation for the excess galaxies detected by JWST at $z>10$. 

\begin{table}
\centering
\caption{Cosmological parameters for the CPL$n\Lambda$CDM model that remains consistent with other cosmological observations and yields the largest enhancement in the abundance of dark matter haloes w.r.t. the baseline \textbf{$\bm{\Lambda}$CDM (Planck 2018)} model, at $z = 13.2$ compared to that at $z = 5$. This model is referred to as the \textbf{relative-zboost CPL$\bm{n\Lambda}$CDM} model in the text.  }
\begin{tabular}{|c|c|c|c|c|c|}
\hline
$\Omega_m$ & $\Omega_b h^2$ & $\Omega_\Lambda$ & $h$ & $\Omega_\phi$ \\
\hline
0.320444 & 0.0220481 & -1.38512 & 0.66883 & 2.06459 \\
\hline
 $w_0$ & $w_a$  & $n_s$ & $\sigma_8$ & $\tau_\mathrm{el}$ \\
\hline
 -0.956524 & -0.165338  & 0.970466 & 0.834091 & 0.0750381 \\
\hline
\end{tabular}
\label{tab:cosmo_params_max_relative-zboost}
\end{table}

\newpage


\bsp	
\label{lastpage}
\end{document}